\documentstyle[12pt,twoside]{article}

\newcommand{\beqn}{\begin{eqnarray}}
\newcommand{\eeqn}{\end{eqnarray}}
\newcommand{\be}{\begin{equation}}
\newcommand{\ee}{\end{equation}}
\newcommand{\ba}{\begin{array}}
\newcommand{\ea}{\end{array}}
\newcommand{\bfr}{\begin{flushright}}
\newcommand{\efr}{\end{flushright}}
\newcommand{\bfl}{\begin{flushleft}}
\newcommand{\efl}{\end{flushleft}}
\newcommand{\bre}{|\kern-.15em|\kern-.15em|}
\newcommand{\br}{|\kern-.25em|\kern-.25em|}
\newcommand{\brr}{{|\kern-.15em|\kern-.15em|\kern-.15em}\,}
\newcommand{\bo}{{\hfill\loota}}
\newcommand{\loota}{\hbox{\enspace{\vrule height 7pt depth 0pt width 7pt}}}

\newcommand{\al}{\alpha}
\newcommand{\C}{{\rm\bf C}}
\newcommand{\ci}{\cite}
\newcommand{\de}{\delta}
\newcommand{\De}{\Delta}
\newcommand{\ds}{\displaystyle}
\newcommand{\fr}{\frac}
\newcommand{\ga}{\gamma}
\newcommand{\HH}{{\bf H}}
\newcommand{\la}{\label}

\newcommand{\na}{\nabla}
\newcommand{\Om}{\Omega}
\newcommand{\om}{\omega}
\newcommand{\ov}{\overline}

\newcommand{\pa}{\partial}

\newcommand{\re}{\ref}

\newcommand{\Si}{\Sigma}
\newcommand{\si}{\sigma}
\newcommand{\ti}{\tilde}
\newcommand{\ve}{\varepsilon}

\def\N{{\rm I\kern-.1567em N}}                              
\def\R{{\rm I\kern-.1567em R}}                              
\def\P{{\rm I\kern-.1567em P}}                              
\def\C{{\rm C\kern-4.7pt                                    
\vrule height 7.7pt width 0.4pt depth -0.5pt \phantom {.}}\,}
\def\Z{{\sf Z\kern-4.5pt Z}}                                

\newcommand{\const}{\mathop{\rm const}\nolimits}

\newcommand{\supp}{\mathop{\rm supp}\nolimits}

\newcommand{\tr}{\mathop{\rm tr\,}\nolimits}

\newtheorem{theorem}{Theorem}[section]

\newtheorem{definition}[theorem]{Definition}

\newtheorem{lemma}[theorem]{Lemma}

\newtheorem{remark}[theorem]{Remark}

\newtheorem{cor}[theorem]{Corollary}
\newtheorem{pro}[theorem]{Proposition}

\begin{document}
\begin{titlepage}
\hspace{6cm} 
{\it Russ. J. Math. Physics} {\bf 12} (2005), no. 3, 301-325  
~\bigskip\bigskip\\
\begin{center}
{\Large\bf On the Convergence to a Statistical Equilibrium \\  
~\\
in the  Crystal  Coupled to a  Scalar Field}  
\end{center}
\vspace{1cm}
 \begin{center}
{\large T.V. Dudnikova}
\footnote{Partially supported by the research grants 
of DFG (436 RUS 113/615/0-1) and 
of RFBR (03-01-00189) and by
the Austrian Science Foundation
(FWF) Project (P16105-N05)}\\
{\small\it M.V. Keldysh Institute
of Applied Mathematics RAS\\
 Moscow 125047, Russia\\
e-mail:~dudnik@elsite.ru}
\bigskip\\
{\large A.I. Komech
\footnote{
On leave from
Department of Mechanics and Mathematics at the
 Moscow State University.
Partially supported by Max-Planck-Institute for Mathematics in the Sciences (Leipzig) and by the  
Wolfgang Pauli Institute (Wien)}}
\\
{\small
\it Fakult\"at der Mathematik, Universit\"at Wien \\
1090 Wien, Austria\\
 e-mail:~komech@mat.univie.ac.at}
\end{center}
 \vspace{2cm}

 \begin{abstract}
We consider the dynamics of a field coupled 
to a harmonic crystal
with $n$ components in  dimension $d$,
 $d,n \ge 1$.
The crystal and the dynamics are
translation-invariant with respect to the subgroup $\Z^d$ of $\R^d$.
 The initial data  is a  random function
with a finite mean density of energy
which also satisfies  a Rosenblatt- or
Ibragimov-Linnik-type mixing condition.
Moreover, initial 
correlation functions are translation-invariant
with respect to the discrete subgroup $\Z^d$.
We study the distribution $\mu_t$
 of the  solution at time $t\in\R$.
The main result is the convergence of $\mu_t$ 
to  a Gaussian  measure as $t\to\infty$,
where $\mu_\infty$ is  translation-invariant
with respect to the subgroup $\Z^d$.

 \end{abstract}
\end{titlepage}

\section{Introduction}

The paper concerns  problems of long-time convergence to an
 equilibrium distribution in a coupled system which is
similar to the Born-Oppenheimer model of a solid state.
In  \cite{DKKS,DKRS,DKS,DKM2} we have started
the  convergence analysis for partial 
differential equations of hyperbolic type in $\R^d$.
In  \ci{DKS1, DKM1} we have extended the results to 
 harmonic crystals.

Here we treat a harmonic crystal coupled
to a scalar Klein-Gordon field.
In this case, the corresponding 
problem in the unit cell is an infinite-dimensional
Schr\"odinger operator, whereas in  
 \ci{DKS1, DKM1} (and in \ci{DKKS,DKRS,DKS,DKM2})
it was a finite-dimensional matrix.
This situation usually arises  in the solid-state
problems similar to that for 
the Schr\"odinger equation with 
space-periodic potential \ci{PST}. The 
main novelty in our methods consists
in that they yield exact estimates of trace norms for 
the problem in the unit cell.

We assume that an initial state $Y_0$ of the coupled system
 is a random element of a Hilbert phase space ${\cal E}$,
see Definition \ref{d1.1}.
  The distribution of $Y_0$ is a probability measure  $\mu_0$
of mean zero satisfying
conditions {\bf S1}-{\bf S3}.
In particular, the measure  $\mu_0$  is invariant
with respect to translations by  vectors of $\Z^d$.
For a given $t\in\R$, we denote by $\mu_t$ the probability measure
defining the distribution of the
 solution $Y(t)$ to the dynamical equations
with  random initial state $Y_0$.
 We study the asymptotics of $\mu_t$ as $t\to\pm\infty$.

Our main result gives  the (weak) convergence of the measures
$\mu_t$  to a limit measure $\mu_{\infty}$,
\be\la{1.8i}
\mu_t \rightharpoondown
\mu_\infty,\,\,\,\, t\to \infty.
\ee
The measure $\mu_{\infty}$ is   Gaussian 
and translation-invariant
with respect to the group $\Z^d$.
We give the explicit formulas for the  covariance
 of  the measure $\mu_\infty$.
The dynamical group is ergodic and mixing with respect to
the limit measure $\mu_\infty$.
Similar results hold as $t\to-\infty$
because the dynamics is time-reversible.

Similar results  have  been established in \ci{BPT, SL}
for  one-dimensional chains of harmonic oscillators
(with $d=1$)
and in \ci{EPR1, FL, JP, RT}
for one-dimensional chains of anharmonic oscillators coupled to heat baths.
For $d$-dimensional harmonic crystals,
with $d\ge 1$, the convergence (\ref{1.8i}) was proved
in \cite{DKS1, DKM1, LL}. 
The mixing condition was  first introduced
 by R.~Dobrushin and Yu.~Suhov for an ideal gas in
 \cite{DS}. The condition
can replace the (quasi-) ergodic hypothesis
when proving the convergence
to the equilibrium distribution, and this 
plays a crucial role in our  approach. 
Developing a Bernstein-type approach,
we have proved the convergence for the wave and Klein-Gordon 
equations and for harmonic crystals
with translation-invariant initial measures, 
 \cite{DKKS,DKRS, DKS1}.
In \cite{DKS, DKM1, DKM2} 
we have extended the results to two-temperature initial measures.
The present paper extends our previous results
to the scalar Klein-Gordon  field 
coupled to the nearest neighbor crystal.

Let us outline our main result and the strategy of the proof.
(For the formal definitions and statements, see Section 2.)
Consider the  Hamilton system with the following
Hamiltonian  functional:
\beqn\la{H}
H(\psi,u,\pi,v)&=&
\frac{1}{2}
\int\Bigr(|\nabla\psi(x)|^2+
|\pi(x)|^2+m_0^2|\psi(x)|^2\Bigr)\,dx
\nonumber\\ 
&+&\frac{1}{2}\sum\limits_{k\in\Z^d}
\Big(\sum\limits_{j=1}^d
|u(k+e_j)-u(k)|^2+|v(k)|^2+\nu_0^2 |u(k)|^2\Big)
\nonumber\\
&+&\sum\limits_{k\in\Z^d}\int
R(x-k)\cdot u(k)\psi(x)\,dx,
\eeqn
involving a real scalar field
$\psi(x)$ and its momentum  $\pi(x)$, $x\in\R^d$,
coupled  to a "simple lattice" described by the
deviations $u(k)\in\R^n$ of the "atoms" and their velocities
$v(k)\in\R^n$, $k\in \Z^d$.
The symbol  $R(x)$ stands for a $\R^n$-valued function
and  $e_j\in\Z^d$ for
 the vector with the coordinates $e_j^i:=\de_j^i$.
Taking the variational derivatives
of $H(\psi,u,\pi,v)$,
we formally obtain the following system
for $x\in\R^d$ and $k\in\Z^d$:
\beqn\la{1}
\left\{\ba{rlrll}
\dot\psi(x,t)&=&\ds\frac{\delta H}{\delta\pi}&=&
\pi(x,t),
\medskip\\
\dot u(k,t)&=&\ds\frac{\partial H}{\partial v}&=&
v(k,t),
\medskip\\
\dot\pi(x,t)&=&\ds-\frac{\delta H}{\delta\psi}&=&
(\De-m_0^2)\psi(x,t)- \sum\limits_{k'\in\Z^d} u(k',t)\cdot R(x-k'),
\medskip\\
\dot v(k,t)&=&\ds-\frac{\partial H}{\partial u}&=&
(\De_L- \nu_0^2)u(k,t)- \ds\int R(x'-k)\psi(x',t)\,dx'.
\ea\right.
\eeqn
Here $m_0,\nu_0> 0$, and $\Delta_L$ denotes for the
discrete Laplace operator on the lattice $\Z^d$,
$$
\Delta_L u(k):=\sum\limits_{e,|e|=1}(u(k+e)-u(k)).
$$
Note that for
$n=d$ and $R(x)=-\nabla \rho(x)$,
the interaction term in the Hamiltonian
is the linearized Pauli-Fierz approximation
of  the translation-invariant coupling
\be\la{1.3'}
\sum_k\ds\int\rho(x-k-u(k))\psi(x)\,dx.
\ee
A similar model was analyzed  by Born and
Oppenheimer \ci{BO}  as a model of a solid state
(coupled Maxwell-Schr\"odinger equations
for  electrons in the harmonic crystal; see, e.g., \ci{PST}).
The traditional analysis of the coupled field-crystal 
system (\ref{1}) is based on an iterative perturbation procedure
 using the {\it adiabatic approximation}. Namely, in the zero 
approximation, the crystal and the (electron) field are 
discoupled.
In the first one, the electron field defines 
a slow displacement of nuclei.
The displacements give the corresponding contribution to the field via 
the static Coulombic potentials, which means
 a non-relativistic approximation, etc.
The iterations converge if the motion of the nuclei is
{\it sufficiently slow}, i.e.,
 {\it the nuclei are rather heavy
as compared with the electrons}.
A similar procedure applies to the corresponding stationary problem of finding the dispersion relations.

Our analysis of the dispersion relations is a bit different
and holds for {\it  small displacements}.
Namely, we linearize the translation-invariant coupling (\ref{1.3'}) 
at the zero displacements of the nuclei and obtain the equations (\ref{1}) 
corresponding to the Pauli-Fierz approximation.
On the other hand, we analyze the dispersion relations of the 
linearized equations without any adiabatic or
non-relativistic approximation.
We give an exact nonperturbative
spectral analysis of the coupled system (\ref{1}).
\medskip

We study the Cauchy problem for the system (\ref{1})
with the initial data
\be\la{5}
\left\{\ba{rlrll}
\psi(x,0)&=\psi_0(x),&~~\pi(x,0)&=\pi_0(x),&~~x\in\R^d,\\
u(k,0)&=u_0(k),&~~v(k,0)&=v_0(k),&~~k\in\Z^d.
\ea\right.
\ee
Let  us write 
\beqn\la{1.4'}
&\psi^0:=\psi,\,\,\,\,\psi^1:=\pi,\,\,\,\,u^0:=u,\,\,\,\
u^1:=v,\nonumber\\
&Y(t):=(Y^0(t),Y^1(t))\,\,\,\left|
\ba{c}
 Y^0(t):=(\psi^0(x,t),u^0(k,t)):=(\psi(x,t),u(k,t)),\\
Y^1(t):=(\psi^1(x,t),u^1(k,t)):=(\pi(x,t),v(k,t)).
\ea\right.
\eeqn
In  other words,
 $Y(\cdot,t)$ are functions defined on the 
disjoint union $\P:=\R^d\cup\Z^d$,
\beqn
Y^i(t)= Y^i(p,t):=
\left\{\ba{ll}
\psi^i(x,t),\,\,\,p=x\in\R^d,\\
u^i(k,t),\,\,\,p=k\in\Z^d,\nonumber
\ea\right. \quad i=0,1.
\eeqn
In this case, the system (\ref{1}), (\ref{5}) becomes
  a dynamical problem of the form
\be\la{1.1'}
\dot Y(t)={\cal A}Y(t),\,\,\,t\in\R;\,\,\,\,Y(0)=Y_0.
\ee
Here $Y_0=(\psi_0,u_0,\pi_0,v_0)$ and
\be\la{A}
{\cal A}=J\nabla H(Y)=
\left(\ba{cc}
0 & 1  \\
-{\cal H} & 0\ea\right),\,\,\,\,
{\cal H}=\left(\ba{cc}
-\De+m_0^2 &  S \\
S^*& -\De_L+\nu_0^2
\ea\right),\,\,\,\,
J=\left(\ba{cc} 0 & 1  \\
-1 & 0\ea\right),
\ee
where $S u(x)=\sum_{k\in\Z^d}R(x-k)u(k)$,
$S^* \psi(k)=\ds\int_{\R^d}R(x-k)\psi(x)\,dx$, and
$$
\langle\psi, S u \rangle_{L^2(\R^d)}
=\langle S^*\psi,u\rangle_{[l^2(\Z^d)]^n},\,\,\,\,\psi\in L^2(\R^d),
\,\,\, u\in [l^2(\Z^d)]^n.
$$
We assume that the 
initial datum $Y_0$ is a random function, and the 
initial correlation matrix
$$
Q_0(p,p'):= E\Big(Y_0(p)\otimes Y_0(p') \Big),\,\,\,p,p'\in \P,
$$
 is translation invariant
with respect to translations by $\Z^d$, i.e., 
\be\la{cfti}
Q_0(p+k,p'+k)=Q_0(p,p'),\,\,\,\, p,p'\in \P,
\ee
for any  $k\in\Z^d$.
We also assume that the
initial mean energy densities  are  uniformly bounded,
\beqn
\la{med}
e_F(x)&:=&E(|\nabla\psi_0(x)|^2+
|\psi_0(x)|^2+|\pi_0(x)|^2)\le \bar e_F  <\infty,\quad \mbox{a.a. } x\in \R^d,\\
\la{med0}
e_L&:=& E(|u_0(k)|^2+|v_0(k)|^2)
<\infty,\quad k\in \Z^d.
\eeqn
Finally, we assume  that the measure $\mu_0$ satisfies
a mixing condition
of a Rosenblatt- or Ibragimov-Linnik type, which means that
\be\la{mix}
Y_0(p)\,\,\,\,   and \, \, \,\,Y_0(p')
\,\,\,\,  are\,\,\,\, asymptotically\,\,\,\, independent\,\, \,\,
 as \,\, \,\,
|p-p'|\to\infty.
\ee
Our main result gives the (weak)
convergence (\ref{1.8i}) of $\mu_t$ to a limit measure $\mu_\infty$,
which is a stationary Gaussian probability measure.
\medskip

Let us comment on the methods of the proof.
The key role in our proof is played by the standard
reduction of system (\ref{1.1'})
to the Bloch problem on the torus. 
Namely, we  split $x\in\R^d$ in the form $x=k+y$,
$k\in\Z^d$, $y\in K_1^d:=[0,1]^d$, and apply the Fourier transform
$F_{k\to\theta}$ to the solution
$Y(k,t):=\Big(\psi(k+y,t),u(k,t),\pi(k+y,t),v(k,t)\Big)$,
\beqn\nonumber
\ti Y(\theta,t):=F_{k\to \theta}Y(k,t)\equiv
\sum\limits_{k\in \Z^d}
e^{ik\theta}Y(k,t)
=(\ti\psi(\theta,y,t),\ti u(\theta,t),
\ti\pi(\theta,y,t),\ti v(\theta,t)),\,\,\,\,
\theta\in \R^d,
\eeqn
which is a version of the  Bloch-Floquet transform.
The functions $\ti \psi$,  $\ti \pi$
are periodic with respect to
 $\theta$ and quasi-periodic with respect to $y$,
i.e.,
$$
\ti\psi(\theta,y+m,t)=e^{-im\theta}\psi(\theta,y,t),\,\,\,\,
\ti \pi(\theta,y+m,t)=e^{-im\theta}\pi(\theta,y,t),\,\,\,\,m\in\Z^d.
$$
Further, introduce the Zak transform of $Y(\cdot,t)$ 
(which is 
also known as Lifshitz-Gelfand-Zak transform) (cf \ci[p.5]{PST}) as
\be\la{YPi}
{\cal Z}Y(\cdot,t)\equiv \ti Y_\Pi(\theta,t):=
(\ti\psi_\Pi(\theta,y,t),\ti u(\theta,t),\ti\pi_\Pi(\theta,y,t),\ti v(\theta,t)),
\ee
where
$\ti\psi_\Pi(\theta,y,t):=
e^{iy\theta}\ti\psi(\theta,y,t)$ and
$\ti\pi_\Pi(\theta,y,t):=
e^{iy\theta}\ti\pi(\theta,y,t)$
are  periodic functions with respect to $y$
(and quasi-periodic with respect to $\theta$).
Denote by $T_1^d:=\R^d/\Z^d$
the real  $d$-torus and write ${\cal R}:=T_1^d\cup\{0\}$.
Set
\beqn\nonumber
\ti Y_\Pi(\theta,r,t)\equiv 
\ti Y_\Pi(\theta,t):=\left\{
\ba{lll}
(\ti\psi_\Pi(\theta,y,t),\ti \pi_\Pi(\theta,y,t)),&r=y\in T_1^d,\\
(\ti u(\theta,t),\ti v(\theta,t)),&r=0.
\ea
\right.
\eeqn
Problem (\ref{1.1'}) is now equivalent to the problem  
on the unit torus $y\in T_1^d$
with the parameter $\theta\in K^d\equiv [0,2\pi]^d$,
\beqn\la{CPF}
\left\{\ba{ll}
\dot{\ti Y}_\Pi(\theta,t)=
\ti {\cal A}(\theta)\ti Y_\Pi(\theta,t),\,\,\,t\in\R\\
\,\,\,\,\ti Y_\Pi(\theta,0)=\ti Y_{0\Pi}(\theta)
\ea
\right|\,\,\,\,\theta\in K^d.
\eeqn
Here
\be\la{tiA}
\ti{\cal A}(\theta)=\left(
 \ba{cc}
0 & 1\\
-\ti{\cal H}(\theta) & 0
\ea\right),
\ee
and $\ti{\cal H}(\theta):={\cal Z}
{\cal H}{\cal Z}^{-1}$
is the ``Schr\"odinger operator'' 
on the torus $T_1^d$,
\be\la{H(theta)}
\ti {\cal H}(\theta)=\left(
\ba{cc}
(i\nabla_y+\theta)^2+m_0^2 & \ti S(\theta)\\
\ti S^*(\theta) & \omega_*^2(\theta)
\ea\right),
\ee
where
\be\la{omega}
\omega_*^2(\theta):=2(1-\cos\theta_1)+\dots+
2(1-\cos\theta_d)+\nu_0^2,
\ee
\be\la{2.6'}
\Big(\ti S(\theta)\ti u(\cdot)\Big)(\theta,y):
=\ti R_\Pi(\theta,y)\cdot\ti u(\theta),\,\,\,
\Big(\ti S^*(\theta) \ti\psi_\Pi(\theta,\cdot)\Big)(\theta)
:=\ds\int_{T_1^d}\ti R_\Pi(-\theta,y)
\ti \psi_\Pi(\theta,y)\,dy,
\ee
$$
\langle
\ti\psi_\Pi(\theta,\cdot), 
(\ti S(\theta) \ti u)(\theta,\cdot) \rangle_{L^2(T_1^d)}
= (\ti S^*(\theta) \ti\psi_\Pi)(\theta)\cdot
\ti u(\theta),\,\,\,\,
\ti\psi_\Pi(\theta,\cdot)\in H^1(T_1^d),
\,\,\, \ti u(\theta)\in \C^n.
$$
Then, formally,
\be\la{solFtr}
\ti Y_\Pi(\theta,t)=
e^{\ti{\cal A}(\theta)t}\ti Y_{0\Pi}(\theta),
\,\,\,\,\theta\in K^d.
\ee
To justify the definition of the exponential,
we note that
$\ti{\cal H}(\theta)$ is a self-adjoint operator with a discrete spectrum.
Indeed, if $R=0$, then this  follows
from  elliptic theory,
and,  if $R\not=0$, then the operators $\ti S(\theta)$
and $\ti S^*(\theta)$ are finite-dimensional
for a fixed $\theta$.
We assume that $\ti{\cal H}(\theta)>0$
(condition {\bf R2})
which corresponds to the hyperbolicity
of problem (\ref{1}).

Note that in  \ci{DKS1, DKM1},
we considered the harmonic crystal without any field.
In this case, the operator $\ti{\cal A}(\theta)$
is a finite-dimensional matrix.

Let us prove the convergence (\re{1.8i}) by using the strategy of
\ci{DKKS}-\ci{DKM2}
in the following three steps.\\
{\bf I.} The family of measures
 $\mu_t$, $t\geq 0$, is weakly
compact in an appropriate Fr\'echet space.\\
{\bf II.} The correlation functions converge to a limit,
 \be\la{corf}
Q_t(p,p'):=\int \Big(Y(p)\otimes Y(p')\Big)\, \mu_t(dY)
\to Q_\infty(p,p'),\,\,\,\,t\to\infty,\,\,\,\,\,p,p'\in \P.
\ee
{\bf III.}
The characteristic functionals converge to a Gaussian 
functional,
\be\la{2.6i}
 \hat\mu_t(Z): =
 \int e^{i\langle Y,Z\rangle }\,\mu_t(dY)
\rightarrow \ds \exp\Big\{-\fr{1}{2}{\cal Q}_\infty (Z,Z)\Big\},
\,\,\,\,t\to\infty,
\ee
where $Z$ is an arbitrary element of the  dual space  and 
${\cal Q}_\infty$ is a quadratic form.

Property {\bf I} follows from the Prokhorov Theorem.
 First, let us prove the  uniform bound (\ref{boundlocenergy})
for the mean local energy in $\mu_t$.
To this end,  we shall show that the operator
$\Big(\Om^i\ti q_t^{ij}(\theta)\Om^j\Big)_{i,j=0,1}$
is of trace class,
where $\ti q_t^{ij}(\theta)$ represents the covariance
of the measure $\mu_t$ in the Zak transform
(see (\ref{qt}))
and $\Om\equiv\Om(\theta):=\sqrt{\ti {\cal H}(\theta)}$.
Moreover,  we derive the uniform bound 
\be\la{traceqt}
\sup_{t\ge 0}\sup_{\theta\in [0,2\pi]^d}
{\rm tr}\Big(\Om^i\ti q_t^{ij}(\theta)\Om^j\Big)_{i,j=0,1}<\infty.
\ee
This implies  the compactness of $\mu_t$ by
 the Prokhorov theorem (when applying
 Sobolev's embedding theorem as in \ci{DKKS}).

To derive property {\bf II},
we study oscillatory integrals  in the Zak transform by
developing our {\it cutting strategy} intoduced in
 \ci{DKS1}.
Namely, we rewrite (\ref{corf}) in the  form
\be\la{corfpsi}
{\cal Q}_t(Z,Z)\to 
{\cal Q}_\infty(Z,Z),\,\,\,t\to\infty,
\ee
where ${\cal Q}_t(Z,Z)$ stands for
 the correlation quadratic form for the measure 
$\mu_t$. Further, we prove formula
(\ref{corfpsi}) for $Z\in{\cal D}^0$ as follows:
by the definition of ${\cal D}^0$,
the  Zak transform $\ti Z_\Pi(\theta)$
vanishes in a neighborhood of a ``critical set''
${\cal C}\subset K^d$.
In particular, 
the set ${\cal C}$ includes  all points $\theta\in K^d$
with a degenerate Hessian  of $\om_l(\theta)$
and the points for which the function
$\om_l(\theta)$ is non-smooth.
One can cut off the critical set ${\cal C}$
  by the following  two crucial observations:
(i) mes${\cal C}=0$ and (ii) 
the initial correlation quadratic form is continuous in $L^2$
due to the mixing condition.
The continuity follows from the spatial decay of the
correlation
functions in accordance with the well-known Shur lemma.

Similarly, we first  prove  property {\bf III}
for $Z\in{\cal D}^0$
and then extend it to all $Z\in{\cal D}$.
For $Z\in{\cal D}^0$, we use a version
of the S.N.~Bernstein ``room-corridor'' technique
(cf. \ci{DKKS, DKS1}).
This leads to a representation of the solution
as the sum of weakly dependent random variables.
Then (\ref{2.6i}) follows from the Central Limit Theorem 
under a Lindeberg-type condition.

Let us comment on the two main technical novelties of our paper.
The first of them is the bound (\ref{traceqt}), 
which then ensures compactness.
We derive formula (\ref{traceqt}) 
in Section 4 directly from our assumption concerning
 the finiteness  of 
the mean energy density (\ref{med}), (\ref{med0}).
The derivation uses the technique of  trace 
class operators \ci{S}, which enables us to avoid 
additional  continuity conditions for 
higher-order derivatives of the correlation functions.
An essential ingredient of the proof is the 
``unitary trick'' (\ref{trace}), 
which is a natural consequence of the Hamiltonian
structure of  system (\ref{1}).
The second main novelty is the bound (\ref{boundsol})
for the dynamics in weighted norms.
In the Zak transform, the  weighted norms  become
 Sobolev norms with  negative index.
We derive (\ref{boundsol}) in Appendix A,
by using duality arguments,
from the corresponding bounds for the derivatives
of the exponential (\ref{solFtr}).
The bounds for the derivatives follow by 
differentiating the dynamical equations.

Let us comment on
our conditions {\bf E1} and {\bf E2}. The conditions  
are natural generalizations of similar conditions
in \ci{DKS1, DKM1}.
Condition {\bf E1} enables us to apply the stationary phase
method to the oscillatory integral representation for the covariance.
It provides that the stationary points
of the phase functions are non-degenerate.

The paper is organized as follows.
In Section 2 we formally state  our main result.
The compactness (Property {\bf I}) is established in Section 4, the
convergence (\ref{corf}) in Section 6, and the
convergence (\ref{2.6i})
in Section 7. 
In Section 8, mixing properties for the limit measures are proved.
Appendix A concerns the dynamics in the Fourier transform,
in Appendix B we analyze the crossing points
of the dispersion relations, and  in Appendix C
we discuss  the covariance
in the spectral representation.

\setcounter{equation}{0}
\section{Main results}
\subsection{Notation}
We assume that the initial data $Y_0$
are given by an element of the real phase space ${\cal E}$ defined below.
\begin{definition}
Let $H^{s,\al}=H^{s,\al}(\R^d)$, $s\in\R$, $\al\in\R$,
be the  Hilbert space of  distributions
$\psi\in S'(\R^d)$ with  finite norm
$$\Vert \psi\Vert_{s,\al}\equiv
\Vert \langle x\rangle^{\al}\Lambda^s
\psi\Vert_{L^2(\R^d)}<\infty.
$$
For $\psi\in D\equiv C_0^\infty(\R^d)$, write
$F\psi ( \xi)= \ds\int e^{i \xi\cdot x} \psi(x) dx.$
Let
 $\Lambda^s \psi:=F^{-1}_{\xi\to x}(\langle  \xi\rangle^s\hat \psi(\xi))$ and
$\langle  x\rangle:=\sqrt{|x|^2+1}$, where
 $\hat \psi:=F \psi$ stands for the Fourier
transform of a tempered distribution $\psi$.
  \end{definition}
\begin{remark}                 \la{d1.1'}
  For $s=0,1,2,\dots$, the space $ H^{s,\al}(\R^d)$  
 is the  Hilbert space of  real-valued functions $\psi(x)$
   with  finite norm
$$ 
\sum_{|\gamma|\le s}\int
(1+|x|^2)^\al|{\cal D}^\gamma\psi(x)|^2\, dx<\infty,
$$
which is equivalent to $\Vert \psi\Vert^2_{s,\al}$.
  \end{remark}
\begin{definition}                 \la{d1.1''}
 Let $ L^\al$, $\al\in\R$,
 be the  Hilbert space  of
 vector functions  $u(k)\in\R^n$, $k\in\Z^d$,
  with  finite norm
$$ 
\Vert u\Vert^2_{\al} \equiv
 \sum\limits_{k\in\Z^d}(1+|k|^2)^\al\vert u(k)\vert^2
  <\infty.
$$
  \end{definition}
\begin{definition}\la{d1.1}
Let ${\cal E}^{s,\al}:=
 H^{1+s,\al}(\R^d)\oplus L^{\al}\oplus
H^{s,\al}(\R^d)\oplus L^\al$
be the Hilbert space  of vectors $Y\equiv(\psi,u,\pi,v)$
 with  finite norm 
$$
\bre Y\bre^2_{s,\al}=
\Vert\psi\Vert^2_{1+s,\al}+\Vert u \Vert^2_{\al}+
\Vert\pi\Vert^2_{s,\al}+\Vert v \Vert^2_{\al}.
$$
\end{definition}

Choose some $\al$, $\al<-d/2$.
 Assume that $Y_0\in {\cal E}:={\cal E}^{0,\al}.$

Using the standard technique of pseudo-differential operators
and Sobolev's Theorem
(see, e.g., \ci{H3}), one can prove that
 ${\cal E}^{0,\al}={\cal E} \subset {\cal E} ^{s,\beta}$
for every $s<0$ and $\beta<\al$,
and the embedding  is compact.
\begin{definition}
The phase space of  problem
(\ref{1.1'}) is  ${\cal E}:=
{\cal E}^{0,\al}$, $\al<-d/2$.
\end{definition}

Introduce the space $H_1^s:=H^s(T^d_1)\oplus\C^n$, $s\in\R$, 
where $H^s(T^d_1)$ stands for the Sobolev space.
\medskip

We assume that
the following conditions hold
for the real-valued coupling
vector function $R(x)$:\\
{\bf R1.} $R\in C^{\infty}(\R^d)$ and
 $|R(x)|\le \bar R \exp(-\ve |x|)$ with some $\ve>0$
and some $\bar R<\infty$.\\
{\bf R2.} The operator $\ti{\cal H}(\theta)$ is
positive definite for
$\theta\in K^d\equiv[0,2\pi]^d$. This is equivalent
to the uniform bound
\beqn\la{lowerbound}
( X^0,\ti{\cal H}(\theta)X^0)
\ge \kappa^2\Vert X^0\Vert^2_{H^1_1}\,\,\,\, 
\mbox{for }\, X^0\in H^1_1,\,\,\,\,\theta\in K^d,
\eeqn
where $\kappa>0$ is a constant and $(\cdot,\cdot)$
stands for the inner product in $H_1^0$ (see (\ref{H_1^0})).
\begin{remark}
i) Condition {\bf R2} ensures that the operator
$i\ti{\cal A}(\theta)$ is self-adjoint
with respect to the energy inner product.
 This corresponds to the
hyperbolicity of problem (\ref{1}).

ii) Condition {\bf R2} holds, in particular, 
if the following condition
 {\bf R2'} holds (see Remark \ref{rR'}):\\
 {\bf R2'.} $\ds\int_{[0,1]^d}\Big|\sum_{k\in\Z^d}
R(k+y)\Big|^2\,dy< \nu^2_0m_0^2/2$.

iii) Condition  {\bf R2'} holds for functions $R$
satisfying condition {\bf R1} with $\bar R \ve^{-d}\ll 1$.
\end{remark}
 \begin{pro}    \la{p1.1}
Let conditions {\bf R1} and  {\bf R2} hold.
Then 
(i) for any $Y_0 \in {\cal E} $,
 there exists  a unique solution
$Y(t)\in C(\R, {\cal E})$
 to the Cauchy problem (\re{1.1'}).\\
 (ii) The operator $W(t):Y_0\mapsto  Y(t)$
 is continuous in ${\cal E} $ for any $t\in \R$,
\be\la{boundsol}
\sup_{|t|\le T}\bre W(t)Y_0\bre_{0,\al}\le C(T)
\bre Y_0\bre_{0,\al}
\ee
if $\al$ is even and $\al\le -2$.
\end{pro}
{\bf Proof}. {\it (i) Local existence.}
Introduce the matrices
\be\la{A0}
{\cal A}_0:=\left(\ba{cc}
0 & 1  \\
-{\cal H}_0 & 0\ea\right),\,\,\,\,
{\cal H}_0=\left( \ba{cc}
-\De+m_0^2 &  0 \\
0& -\De_L+\nu_0^2
\ea\right).
\ee
Then  problem (\ref{1}) can be rewritten
as  the Duhamel integral 
$$
Y(t)=e^{{\cal A}_0 t}Y_0+\int\limits_0^t
e^{{\cal A}_0 (t-s)}BY(s)ds,
$$
where
$$
BY=\Big(0,0,-\sum\limits_{k'\in\Z^d}u(k')R(x-k'),
-\int\limits_{\R^d}
 R(x'-k)\psi(x')\,dx'\Big),\,\,\,\, Y=(\psi,u,\pi,v).
$$
Condition {\bf R1} implies that
$
\brr BY(s)\brr_{0,\al}\le
C\brr Y(s)\brr_{0,\al}.
$
Further, for $0\le s\le t\le T$, we obtain 
$$
\brr e^{{\cal A}_0 (t-s)}BY(s)\brr_{0,\al}\le
C(T)\brr Y(s)\brr_{0,\al},
$$
(see, e.g., \ci{DKS1}). Hence,
$$
\max_{|t|\le T}\brr Y(t)\brr_{0,\al}\le
C(T)\brr Y_0\brr_{0,\al}+
TC(T)\max_{|s|\le T}\brr Y(s)\brr_{0,\al}
\le(T+1)C(T)\max_{|s|\le T}\brr Y(s)\brr_{0,\al}.
$$
We choose a $T>0$ so that  $(T+1)C(T)<1$.
Then the contraction mapping principle
implies the existence of a unique solution
$Y(t)\in C([0,T];{\cal E})$.
{\it  The global existence} follows from
the bound (\ref{boundsol}).
 
{\it (ii) } The bounds (\ref{boundsol}) are proved in
 Corollary \ref{adjproblem}. \bo\\
\medskip

Conditions {\bf R1} and {\bf R2}
imply that, for a fixed $\theta\in K^d$,
the operator $\ti{\cal H}(\theta)$
is positive definite   and self-adjoint
in $H_1^0$ and its spectrum is discrete.
Introduce the Hermitian positive-definite operator
$$
\Om(\theta):=\sqrt{\ti {\cal H}(\theta)}> 0.
$$
Denote by $\om_l(\theta)> 0$ and
 $F_l(\theta,\cdot)$, $l=1,2,\dots,$ the eigenvalues
(``Bloch bands'') and the  orthonormal eigenvectors
(``Bloch functions'') of the operator
$\Om(\theta)$ in $H_1^0$, respectively.
Note that $F_l(\theta,\cdot)
\in H_1^\infty:=C^\infty(T_1^d)\oplus \C^n$,
because these are eigenfunctions of the elliptic operator
$\ti{\cal H}(\theta)$.

As is well known,  the functions
$\om_l(\cdot)$ and  $F_l(\cdot,r)$ are real-analytic
outside the set of the ``crossing'' points $\theta_*$, where
$\om_l(\theta_*)=\om_{l'}(\theta_*)$ for some  $l\ne l'$.
However,  the functions  are not smooth
at the crossing points in general
 if $\om_l(\theta)\not\equiv\om_{l'}(\theta)$.
Therefore, we need the following lemma, which is proved in
Appendix B.
\begin{lemma}\la{lc*}
(cf. \ci{Wilcox})
There exists
a closed subset ${\cal C}_*\subset K^d$ such that
(i) the Lebesgue measure of ${\cal C}_*$ vanishes,
\be\la{c*}
{\rm mes}\,{\cal C}_*=0.
 \ee
(ii) For every point $\Theta\in K^d\setminus{\cal C}_*$
and $N\in\N$,
there exists a neighborhood
${\cal O}(\Theta)\subset K^d\setminus{\cal C}_*$ such that each of the functions
$\om_l(\theta)$ and  $F_l(\theta,\cdot)$,  $l=1,\dots, N$,
can be chosen to be  real-analytic on
${\cal O}(\Theta)$.\\
(iii) The eigenvalues $\om_l(\theta)$ have constant multiplicity
in ${\cal O}(\Theta)$, i.e., one can enumerate  them
in such a way that 
\beqn
&&\om_1(\theta)\!\equiv\!\dots\!\equiv\!\om_{r_1}(\theta),\,\,\,
\om_{r_1+1}(\theta)\!\equiv\!\dots\equiv\!\om_{r_2}
(\theta),\dots, \la{enum}\\
&&\om_{r_\si}(\theta)\!\not\equiv\!
\om_{r_\nu}(\theta)\,\,\,\,{\rm if}\,\,\,\,\si\ne\nu,
\,\,\, r_\si,r_\nu\ge 1,
\la{enum'}
\eeqn
 for  any $\theta\in {\cal O}(\Theta)$.
\end{lemma}
\begin{cor}\la{civ}
The spectral decomposition holds,
\be\la{spd'}
\Om(\theta)=\sum_{l=1}^{+\infty} \om_l
(\theta)P_l(\theta),
\,\,\,\,\theta\in {\cal O}(\Theta),
\ee
where  $P_l(\theta)$ are the orthogonal projectors
in $H^0_1$ onto the linear span of $F_l(\theta,\cdot)$,
and $P_l(\theta)$ and $\om_l(\theta)$ depend
   on $\theta\in{\cal O}(\Theta)$ analytically.
\end{cor}

Assume that  system (\ref{1.1'}) satisfies
the following conditions {\bf E1} and {\bf E2}.
For every $\Theta\in K^d\setminus{\cal C}_*$:\\
{\bf E1}
$D_l(\theta)\not\equiv 0$,
$l=1,2,\dots$, where
$D_l(\theta):=\det\Big(
\ds\frac{\pa^2\om_l(\theta)}{\pa \theta_i\pa \theta_j}
\Big)_{i,j=1}^{d}$, $\theta\in {\cal O}(\Theta)$,
and ${\cal O}(\Theta)$ is defined in Lemma \ref{lc*}.
\medskip

Write
$$
{\cal C}_l:=
\bigcup\limits_{\Theta\in K^d\setminus {\cal C}_*}
\{\theta\in {\cal O}(\Theta):
\,D_l(\theta)=0\},\,\,\,\, l=1,2,\dots.
$$
The following lemma is also proved in  Appendix B.
\begin{lemma}\la{lc}
Let  conditions {\bf R1} and {\bf R2} hold. Then
${\rm mes}~{\cal C}_l=0$, $l=1,2,\dots.$
\end{lemma}
{\bf E2}
For each $l\ne l'$,  the identity
$\om_l(\theta)-\om_{l'}(\theta)\!\equiv\!{\rm const}_-$,
$\theta\in {\cal O}(\Theta)$,
cannot hold  for any constant const$_-\ne 0$, and
the identity
$\om_l(\theta)+\om_{l'}(\theta)\!\equiv\!{\rm const}_+$
cannot hold  for any constant const$_+\ne 0$.
\medskip

Condition {\bf E2} could be considerably weakened
(cf. \ci[Remark 2.10,  iii, condition E5']{DKS1}).
Note that  conditions {\bf E1} and {\bf E2}
hold if $R=0$.

Let us show that  conditions {\bf E1} and {\bf E2}
hold for ``almost all'' functions $R$
satisfying  conditions {\bf R1}, {\bf R2}.
More precisely, consider  finitely many
coupling functions $R_1,\dots,R_N$
satisfying  conditions {\bf R1} and {\bf R2}'
and take their linear combinations
$$
R_C(x)=\sum_{s=1}^{N}C_sR_s(x),\,\,\,\, C=(C_1,\dots,C_N)\in \R^N.
$$
For $R_C(x)$,  conditions {\bf R1} and {\bf R2}'
hold if $\Vert C\Vert<\ve$ with a sufficiently small 
$\ve>0$.
Let 
$M_1:=\{C\in B_\ve:\, \mbox{condition } {\bf E1}
\,\,\, \mbox{holds for }\,R_C(x)\}$ and  
$M_2:=\{C\in B_\ve:$ condition {\bf E2} holds for
$R_C(x)\}$, 
where $B_\ve:=\{C\in\R^N:\Vert C\Vert<\ve\}$.
In Appendix B, we prove the
following lemma.
\begin{lemma}\la{l45}
The sets $M_1$ and $M_2$ are 
 dense in some ball $B_\ve$
for a  sufficiently small $\ve>0$.
\end{lemma}

\subsection{Random solution. Convergence to equilibrium}

Let $(\Om,\Si,P)$ be a probability space with  expectation $E$
and let ${\cal B}({\cal E} )$ denote the Borel $\si$-algebra
in ${\cal E} $.
Assume that $Y_0=Y_0(\om,p)$ (see (\re{1.1'}))
is a measurable random function
with values in $({\cal E} ,\,{\cal B}({\cal E} ))$.
In other words,
 the map $(\om,p)\mapsto Y_0(\om,p)$ is a measurable map
$\Om\times \P\to\R^{2+2n}$
 with respect to the (completed) $\si$-algebra
$\Sigma\times{\cal B}(\P)$ and ${\cal B}(\R^{2+2n})$.
Then $Y(t)=W(t) Y_0$ is also a measurable    random function
with values in
$({\cal E} ,{\cal B}({\cal E}))$ owing to Proposition \re{p1.1}.
Denote by $\mu_0(dY_0)$ the
 Borel probability measure in ${\cal E} $
giving the distribution of  $Y_0$.
Without loss of generality,
 we can assume 
$(\Om,\Si,P)=({\cal E} ,{\cal B}({\cal E} ),\mu_0)$
and $Y_0(\om,p)=\om(p)$ for
$\mu_0(d\om)\times dp$-almost all
point $(\om,p)\in{\cal E} \times \P$.
\begin{definition}
The measure $\mu_t$ is a Borel probability measure in ${\cal E} $
 giving the distribution of $Y(t)$,
$$\mu_t(B) = \mu_0(W(-t)B),\,\,\,\,
\forall B\in {\cal B}({\cal E} ),
\,\,\,   t\in \R.
$$
\end{definition}

Our main objective is to prove the weak convergence of the measures $\mu_t$
in the Fr\'echet spaces ${\cal E} ^{s,\beta }$ for each  $s<0$,
 $\beta<\al<-d/2$,
\be\la{1.8}
\mu_t\,\buildrel {\hspace{2mm}{\cal E} ^{s,\beta }}\over
{- \hspace{-2mm} \rightharpoondown }
\, \mu_\infty
\quad{\rm as}\quad t\to \infty,
\ee
where $\mu_\infty$ is a limit measure on
${\cal E}\equiv{\cal E} ^{0,\al}$.
 This is equivalent to the convergence
$$
 \int f(Y)\mu_t(dY)\rightarrow
 \int f(Y)\mu_\infty(dY)\quad{\rm as}\quad t\to \infty
$$
 for any bounded continuous functional $f(Y)$
 on  ${\cal E} ^{s,\beta}$.

Let ${\cal D}=[D_F\oplus D_L]^2$ 
with $D_F\equiv C^\infty_0(\R^d)$, and
let $D_L$ be the set of vector sequences $u(k)\in \R^n$, $k\in\Z^d$,
such that $u(k)=0$ for $k\in\Z^d$ outside a finite set.
For a  probability  measure $\mu$ on  ${\cal E}$,
 denote by $\hat\mu$ the characteristic functional (Fourier transform)
$$
\hat \mu(Z)  \equiv  
\int e^{i\langle Y,Z \rangle }\,\mu(dY),\,\,\,
 Z\in  {\cal D}.
$$
Here $\langle \cdot,\cdot \rangle$ stands for the inner
 product in 
$L^2(\P)\otimes\R^N$ with different  $N=1,2,\dots$,  
\beqn
\langle Y,Z \rangle&:=&\sum_{i=0}^1\langle Y^i,Z^i \rangle,
\,\,\,
Y=(Y^0,Y^1),\,\,\,Z=(Z^0,Z^1),\nonumber\\
\langle Y^i,Z^i \rangle&:=&\int\limits_{P}  Y^i(p)Z^i(p)\,dp\equiv
\int\limits_{\R^d}  \psi^i(x) \xi^i(x)\,dx+
\sum\limits_{k\in\Z^d} u^i(k)\chi^i(k),
\nonumber
\eeqn
where $Y^i=(\psi^i,u^i)$, $Z^i=(\xi^i,\chi^i)$.
A  measure $\mu$ is said to be {\it Gaussian}
 (with zero expectation) if
its characteristic functional has the form
$$
\ds\hat {\mu} (Z) =  \ds \exp\Big\{-\fr{1}{2}
 {\cal Q}(Z,Z)\Big\},\,\,\,Z \in {\cal D},
$$
where ${\cal Q}$ is a  real nonnegative quadratic form in ${\cal D}$.
\begin{definition}\la{dcorf}
The correlation functions of the measure $\mu_t$,
$t\in\R,$ are defined by
\be\la{qd}
Q^{ij}_t(p,p')\equiv E \Big(Y^i(p,t)\otimes Y^j(p',t)\Big),
\,\,\,\,i,j=0,1,\,\,\,\, p,p'\in \P,
\ee
where  $E$ stands for the integral with respect to the measure $\mu_0(dY)$ and 
the convergence of the integral in (\ref{qd}) is understood in the sense 
of distributions, namely,
\be\la{qdp}
\langle Q^{ij}_t(p,p'),Z_1(p)\otimes Z_2(p')\rangle:=E \langle  Y^i(p,t),Z_1(p)\rangle\langle 
Y^j(p',t),Z_2(p')\rangle,
\,\,\,\,Z_1, Z_2\in D_F\oplus D_L.
\ee
\end{definition}

\subsection{Mixing condition}
 Let $O(r)$ be the set of all pairs of open subsets
 $ {\cal A}, {\cal B} \subset \P$ such that
the distance
 $\rho ( {\cal A},\, {\cal B})$  is not less than $r$,
 and let  $\sigma ({\cal A})$ be
the $\sigma $-algebra  in ${\cal E}$ generated by the
linear functionals $Y\mapsto\, \langle Y,Z\rangle $
for which $Z\in  {\cal D}$ and
 $ \supp Z  \subset {\cal A}$.
Define the Ibragimov-Linnik mixing coefficient
of a probability  measure  $\mu_0$ on ${\cal E}$
by the formula (cf. \ci[Definition 17.2.2]{IL})
$$\varphi(r)\equiv
\sup_{({\cal A},{\cal B})\in O(r)} \sup_{
\ba{c} A\in\si({\cal A}),B\in\si({\cal B})\\ \mu_0(B)>0\ea}
\fr{| \mu_0(A\cap B) - \mu_0(A)\mu_0(B)|}{ \mu_0(B)}.
$$
\begin{definition}\la{dmix}
A  measure $\mu_0$ satisfies the strong uniform
Ibragimov-Linnik mixing condition if
$\varphi(r)\to 0$ as $\quad r\to\infty$.
\end{definition}
Below we specify  the rate of the decay
of $\varphi$
(see Condition {\bf S3}).

\subsection{Main theorem}

Assume that the initial measure $\mu_0$
satisfies the following properties {\bf S0}--{\bf S3}:\\
{\bf S0.} The measure
$\mu_0$ has zero expectation value,
$E Y_0(p)  \equiv  0$, $p\in \P$.\\
{\bf S1.} The correlation matrices of $\mu_0$  are
  invariant 
with respect to translations in $\Z^d$, i.e.,
Eqn (\ref{cfti}) holds for a.a.  $p,p'\in \P$.\\
{\bf S2.} The measure $\mu_0$  has  a finite  mean ``energy'' density,
i.e., Eqns (\ref{med}),  (\ref{med0}) hold.\\
{\bf S3.} The measure $\mu_0$ satisfies the {\it strong uniform}
Ibragimov-Linnik mixing condition with
\be\la{1.12}
\int\limits_0^{+\infty} r^{d-1}\varphi^{1/2}(r)dr <\infty.
\ee
Introduce  the correlation matrix $Q_\infty(p,p')$ of the limit
measure $\mu_\infty$.
It is translation-invariant with respect to translations in $\Z^d$,
i.e.,
\be\la{Qinfty}
Q_\infty(p,p')= Q_\infty(p+k,p'+k),\,\,\,\,k\in\Z^d.
\ee
For $Z\in{\cal D}$, write
\beqn\la{Qinfty'}
{\cal Q}_\infty(Z,Z):=\langle Q_{\infty}(p,p'),Z(p)\otimes Z(p')\rangle
=(2\pi)^{-d} \int\limits_{K^d}
\Big(\ti q_\infty (\theta),
\ti Z_\Pi(\theta,\cdot)\otimes\overline{\ti Z_\Pi(\theta,\cdot)}\Big)
d\theta,
\eeqn
where $\ti q_\infty(\theta)$ is the operator-valued function given by the rule
\beqn\la{qinfty+}
\ti q_\infty(\theta):=
 \sum_{l=1}^{+\infty}P_l(\theta)\frac{1}{2}
\left(\ba{cc}
\ti q^{00}_0(\theta)+\ti{\cal H}^{-1}(\theta)\ti q^{11}_0(\theta)&
\ti q^{01}_0(\theta)-\ti q^{10}_0(\theta)\\
\ti q^{10}_0(\theta)-\ti q^{01}_0(\theta)&
\ti {\cal H}(\theta)\ti q^{00}_0(\theta)+\ti q^{11}_0(\theta)\\
\ea \right)P_l(\theta),
\eeqn
for $\theta\in K^d\setminus{\cal C}_*$.
Here the symbol $\ti q^{ij}_0(\theta)={\rm Op}
\Big((\ti q_0^{ij}(\theta,r,r')\Big)$
stands for the integral operator with the integral kernel 
$\ti q^{ij}_0(\theta,r,r')$ (see formula
(\ref{qt}) with $t=0$), $r,r'\in{\cal R}$,
and $P_l(\theta)$ is the spectral projection operator
introduced in Corollary \ref{civ}.
\medskip\\
{\bf Theorem A} {\it
Let  the conditions {\bf S0}--{\bf S3}, {\bf R1}, {\bf R2},
{\bf E1} and {\bf E2} hold.
Then the following assertions are valid.\\
(i) The convergence (\ref{1.8}) holds for any $s<0$ and
 $\beta<-d/2$. \\
(ii)  The limit measure
$ \mu_\infty $ is  Gaussian  on ${\cal E}$.\\
(iii) The  characteristic functional of $ \mu_{\infty}$
is  Gaussian,
$$
\ds\hat { \mu}_\infty (Z) =
\exp\{-\fr{1}{2}  {\cal Q}_\infty  (Z,Z)\},\,\,\,
Z \in {\cal D}.
$$
(iv) The measure $\mu_\infty$ is invariant, i.e.,
$[W(t)]^*\mu_\infty=\mu_\infty$, $t\in\R$.}
\medskip

Assertions {\it (i)-(iii)} of Theorem~A follow  
from Propositions  \re{l2.1} and \re{l2.2} below.
 \begin{pro}\la{l2.1}
  The family of  measures $\{\mu_t,\,t\in \R\}$
 is weakly compact in   ${\cal E}^{s,\beta}$ with any
 $s<0$ and $\beta<\al<-d/2$, and the following
bounds hold:
\be\la{boundlocenergy}
\sup\limits_{t\in \R}
 E \bre W(t)Y_0\bre^2_{0,\al} \le C(\al)<\infty.
\ee
 \end{pro}
 \begin{pro}\la{l2.2}
 The convergence  (\ref{2.6i})
holds for every $Z\in {\cal D}$.
 \end{pro}

Proposition  \ref{l2.1} (Proposition \ref{l2.2})
 provides the existence (the uniqueness)
of the limit measure $\mu_\infty$.
They are proved in Sections 4 and 7, respectively.

Theorem~A {\it (iv)} follows from
 (\re{1.8})  because the group $W(t)$ is continuous 
with respect to
${\cal E}$ by Proposition~\re{p1.1} {\it (ii)}.

\setcounter{equation}{0}
\section{  Correlation matrices }
To prove the compactness of the family of measures $\{\mu_t\}$,
we  introduce auxiliary notations and prove
necessary bounds for initial correlation matrices.
Since $Y^i(p,t)= (\psi^i(x,t),u^i(k,t))$,
we can rewrite formula (\ref{qd}) as follows:
\beqn\la{5.1}
Q_t^{ij}(p,p')&=&E[Y^i(p,t)\otimes Y^j(p',t)]
=\left(\ba{cc}
E\Big(\psi^i(x,t)\otimes \psi^j(x',t) \Big)&
E\Big(\psi^i(x,t)\otimes u^j(k',t) \Big)\\
E\Big(u^i(k,t)\otimes \psi^j(x',t) \Big)&
E\Big(u^i(k,t)\otimes u^j(k',t) \Big)\ea\right)
\nonumber\\
&\equiv&\left(\ba{cc}
Q_t^{\psi^i\psi^j}(x,x') &
Q_t^{\psi^i u^j}(x,k')\\
Q_t^{u^i\psi^j}(k,x')&
Q_t^{u^iu^j}(k,k')\ea\right),\,\,\,\,i,j=0,1,
\,\,\,\,t\in\R.
\eeqn
Let us rewrite the correlation matrices
$Q_t^{ij}(p,p')$ by using the condition {\bf S1}.
Note that the dynamical group 
$W(t)$ commutes with the translations in $\Z^d$.
In this case, condition  {\bf S1} implies that
\be\la{invQt}
Q_t(k+p,k+p')=Q_t(p,p'),\,\,\,\,t\in\R,\,\,\,\,k\in\Z^d.
\ee
Let us introduce
the splitting $p=k+r$, where $k\in \Z^d$ and $r\in K_1^d\cup 0$.
In other words,
\beqn\nonumber
r=\left\{\ba{ll}
x-[x]\in K_1^d,&\mbox{if } p=x\in\R^d,\\
0,&\mbox{if } p=k\in\Z^d.
\ea\right.
\eeqn 
In this notation,  (\ref{invQt}) implies that
\beqn\la{Qtkr}
Q_t^{ij}(k+r,k'+r')=
:q_t^{ij}(k-k',r,r')\equiv
\left(\ba{cc}
q_t^{\psi^i\psi^j}(k-k'+r,r')&
q_t^{\psi^i u^j}(k-k'+r)\\
q_t^{u^i\psi^j}(k'-k+r')&
q_t^{u^iu^j}(k-k')\ea\right).
\eeqn
Using  the Zak transform (\ref{YPi}), 
 introduce the following matrices
(cf (\ref{qdp})):
\beqn\la{qdpFt}
\ti Q_t^{ij}(\theta,r,\theta',r'):
=E[\ti  Y^i_{\Pi}(\theta,r,t)\otimes
\overline{\ti Y^j_{\Pi}(\theta',r',t)}],\,\,\,\,
\theta,\theta'\in K^d,\,\,\,\, r,r'\in{\cal R}\equiv T_1^d\cup 0,
\eeqn
where the convergence of the mathematical expectation  
is understood in the sense of distributions.
Namely, write
 $\ti{\cal D}=[\ti{\cal D}_F\oplus\ti{\cal D}_L]^2 $
and $\ti{\cal D}_F:=C^\infty(K^d\times T_1^d)$,
$\ti{\cal D}_L:=[C^\infty(T^d)]^n$. Then
\beqn
\langle\ti Q_t^{ij}(\theta,r,\theta',r'),
\ti Z^i_\Pi(\theta,r)\otimes
\overline{\ti Z^j_\Pi(\theta',r')}\rangle
=E\langle\ti  Y^i_{\Pi}(\theta,r,t),
\ti Z^i_\Pi(\theta,r)\rangle\,
\overline{
\langle\ti Y^j_{\Pi}(\theta',r',t),
\ti Z^j_\Pi(\theta',r')\rangle}
\eeqn
for $\ti Z_\Pi=(\ti Z^0_\Pi,\ti Z^1_\Pi)\in \ti{\cal D}$. Now
(\ref{invQt}) implies that
\beqn\la{ap22}
\ti Q_t^{ij}(\theta,r,\theta',r')=
(2\pi)^d\delta(\theta-\theta')
\ti q_t^{ij}(\theta,r,r'),\,\,\,\,\theta,\theta'\in K^d,
\,\,\,\,r,r'\in {\cal R},\,\,\,\,t\in\R,
\eeqn
where
\beqn\la{qt}
\ti q_t^{ij}(\theta,r,r')=
e^{i (r-r')\theta}\sum\limits_{k\in\Z^d}
e^{ik\theta}q_t^{ij}(k,r,r')=
\left(\ba{cc}
\ti q_t^{\psi^i\psi^j}(\theta,y,y')&
\ti q_t^{\psi^i u^j}(\theta,y)\\
\overline{\ti q_t^{u^i\psi^j}}(\theta,y')&
\ti q_t^{u^iu^j}(\theta)
\ea\right).
\eeqn

Recall that 
$\P$ is the disjoint union $\R^d\cup\Z^d$.
For a measurable function $Y(p)$, write
\beqn\la{I}
\int\limits_{\P}Y(p)\,dp=
\int\limits_{\R^d}Y(x)\,dx+
\sum\limits_{k\in\Z^d}Y(k).
\eeqn
\begin{pro}\la{l4.1}
Let conditions {\bf S0}--{\bf S3} hold. Then
(i) the following bounds hold
\beqn
\int\limits_{\P} |Q_0(p,p')|\,dp
&\le& C<\infty\,\,\,\mbox{ for any }\,p'\in \P,
\la{pr1}\\
\int\limits_{\P} |Q_0(p,p')|\,dp'
&\le& C<\infty\,\,\,\mbox{ for any }\,p\in \P,
\la{pr2}
\eeqn
where the constant $C$
does not depend on $p,p'\in \P$. \\
(ii)
${\cal D}_{y,y'}^{\al,\beta}
\ti q_0^{\psi^i\psi^j}(\theta,y,y')$,
${\cal D}_{y}^{\al}
\ti q_0^{\psi^i u^j}(\theta,y)$,
${\cal D}_{y'}^{\beta}
\ti q_0^{u^i\psi^j}(\theta,y')$,
$\ti q_0^{u^iu^j}(\theta)$
are uniformly bounded
in $(\theta,y,y')\in K^d\times T_1^d\times T_1^d$,
$|\al|\le1-i$, $|\beta|\le 1-j$.
\end{pro}
{\bf Proof}.
{\it  (i)}
By \ci[Lemma 17.2.3]{IL}, conditions {\bf S0},
{\bf S2} and {\bf S3} imply
\be\la{4.9'}
|Q_0(p,p')|
\le C \max\{\bar e_F,e_L\}\,\varphi^{1/2}(|p-p'|),~~ p,p'\in \P.
\ee
Hence, the bounds
(\ref{pr1}) and (\ref{pr2}) follow from (\ref{1.12}).

{\it  (ii)}
Similarly to (\ref{4.9'}), 
 conditions {\bf S0}, {\bf S2} and {\bf S3} imply
\beqn
|{\cal D}_{y,y'}^{\al,\beta}
\ti q_0^{\psi^i\psi^j}(\theta,y,y')|&\le&
\sum\limits_{k\in\Z^d}|{\cal D}_{y,y'}^{\al,\beta}
 q_0^{\psi^i\psi^j}(k+y,y')|\le
C\sum\limits_{k\in\Z^d}\varphi^{1/2}(|k+y-y'|)\nonumber\\
&\le&
C\sum\limits_{k\in\Z^d}\varphi^{1/2}(|k|-2\sqrt d)
\le C<\infty.\nonumber
\eeqn
Similar arguments imply the other bounds.
\hfill$\bo$
\begin{cor}\la{c4.1}
By the Shur lemma, Proposition \ref{l4.1}, (i)
implies  that the following bound holds
for any $F,G\in {\bf L}^2:=[L^2(\P,dp)]^2=
 [L^2(\R^d)\oplus [l^2(\Z^d)]^n]^2$:
\beqn\nonumber
|\langle
Q_0(p,p'),F(p)\otimes G(p')\rangle|\le
C\Vert F\Vert_{{\bf L}^2}
\Vert G\Vert_{{\bf L}^2}.
\eeqn
\end{cor}
\begin{cor}\la{c4.2}
The quadratic form ${\cal Q}_\infty(Z,Z)$
defined in (\ref{Qinfty'})--(\ref{qinfty+})
is continuous in ${\bf L}^2$.
\end{cor}
{\bf Proof}. Formulas (\ref{Qinfty'}) and
(\ref{qinfty+}) imply
\beqn\la{310'}
&&\langle Q_\infty(p,p'), Z(p)\otimes Z(p')\rangle=
(2\pi)^{-d}
\int\limits_{K^d\setminus{\cal C}_*}
\Big(\ti q_\infty(\theta,r,r'),
\ti Z_\Pi(\theta,r)\otimes
\overline{\ti Z_\Pi(\theta,r')}\Big)\,d\theta
\nonumber\\
&=&(2\pi)^{-d}\frac{1}{2}
\sum\limits_{l=1}^{\infty}
\int\limits_{K^d\setminus{\cal C}_*}\Big(
\ti q_0(\theta,r,r')+\ti r_0(\theta,r,r'),
P_l(\theta)\ti Z_\Pi(\theta,r)
\otimes P_l(\theta)\overline{\ti Z_\Pi(\theta,r')}\Big)\,d\theta,\,\,\,\,\,
\eeqn
where $\ti r_0(\theta,r,r')$ is the integral kernel
of the operator 
$\ti r_0(\theta):=
\left(\ba{cc}
\ti{\cal H}^{-1}(\theta)\ti q^{11}_0(\theta)&
-\ti q^{10}_0(\theta)\\
-\ti q^{01}_0(\theta)&\ti {\cal H}(\theta)\ti q^{00}_0(\theta)
\ea\right)$.
Here and below, the symbol  $(\cdot,\cdot)$
stands for the inner product in $H_1^0\equiv
H^0(K^d_1)\oplus\C^n$, i.e.,
\be\la{H_1^0}
(F,G)=\int\limits_{K_1^d}\overline{ F^1}(y) G^1(y)\,dy+
\overline{F^2} \cdot G^2,\,\,\,\,\,F=(F^1,F^2),
G=(G^1,G^2)\in H_1^0,
\ee
or in $\HH^0\equiv [H_1^0]^2$.
Consider the terms in the RHS of (\ref{310'}).
Since
\be\la{310''}
\sum\limits_{l=1}^{\infty}
\Vert P_l \ti Z_\Pi \Vert^2_{[L^2(K^d\times {\cal R})]^2}
\le C\Vert\ti Z_\Pi \Vert^2_{[L^2(K^d\times{\cal R})]^2}
=C' \Vert Z\Vert^2_{{\bf L}^2},
\ee
we obtain
\be\la{311}
\sum\limits_{l=1}^{\infty}
\int\limits_{K^d\setminus{\cal C}_*}\Big( \ti q^{ij}_0(\theta,r,r'),
P_l(\theta)\ti Z^{i'}_\Pi(\theta,r)
\otimes P_l(\theta)\overline{\ti Z^{j'}_\Pi(\theta,r')}\Big)
\,d\theta
\le C\Vert Z\Vert^2_{{\bf L}^2},\,\,\,i,j,i',j'=0,1,
\ee
by Corollary \ref{c4.1}.
Further, consider $\ti r_0^{ij}$.
For the terms with $\ti r_0^{01}$ and $\ti r_0^{10}$,
  estimate (\ref{311}) holds.
We rewrite the term with $\ti r_0^{00}$ in the form
\beqn\la{312}
\sum\limits_{l=1}^{\infty}
\int\limits_{K^d\setminus{\cal C}_*}
\Big(\ti r^{00}_0(\theta,r,r'),
P_l(\theta)\ti Z^0_\Pi(\theta,r)
\otimes P_l(\theta)\overline{\ti Z^{0}_\Pi(\theta,r')}\Big)\,
d\theta
\nonumber\\
=\sum\limits_{l=1}^{\infty}
\int\limits_{K^d\setminus{\cal C}_*}\Big( 
 \ti q^{11}_0(\theta,r,r'),
\ti{\cal H}^{-1}(\theta)P_l(\theta)\ti Z^0_\Pi(\theta,r)
\otimes P_l(\theta)\overline{\ti Z^{0}_\Pi(\theta,r')}\Big)\,d\theta.
\eeqn
It follows from  estimates  (\ref{310''}) and (\ref{estH-1})
and  Corollary  \ref{c4.1}  that
the RHS of (\ref{312}) is estimated from above by
$C\Vert Z^0\Vert^2_{L^2(\P)}$.
Finally, consider the term with $\ti r_0^{11}$
and represent it in the form
\beqn\la{313}
\sum\limits_{l=1}^{\infty}
\int\limits_{K^d\setminus{\cal C}_*}\Big( 
 \ti r^{11}_0(\theta,r,r'),
P_l(\theta)\ti Z^1_\Pi(\theta,r)
\otimes P_l(\theta)\overline{\ti Z^{1}_\Pi(\theta,r')}\Big)\,d\theta
\nonumber\\
=\sum\limits_{l=1}^{\infty}
\int\limits_{K^d\setminus{\cal C}_*}
\Big(  \ti q^{00}_0(\theta,r,r'),
\Om(\theta)P_l(\theta)\ti Z^1_\Pi(\theta,r)
\otimes \Om(\theta)
P_l(\theta)\overline{\ti Z^{1}_\Pi(\theta,r')}\Big)\,d\theta.
\eeqn
By the  bounds (\ref{310''}) and (\ref{estOm}),
the RHS of (\ref{313}) is bounded by
\beqn\la{314}
C\int\limits_{K^d}\Big(\Vert \ti q_0^{\psi^0\psi^0}
(\theta,\cdot,\cdot)\Vert_{[H^1(T^d_1)]^2}+
\Vert \ti q_0^{\psi^0 u^0}(\theta,\cdot)\Vert_{H^1(T^d_1)}
+\Vert \ti q_0^{u^0\psi^0 }(\theta,\cdot)\Vert_{H^1(T^d_1)}
\nonumber\\
+|\ti q_0^{u^0 u^0 }(\theta)|\Big)
\Vert \ti Z^1_\Pi(\theta,\cdot)\Vert^2_{H^0_1}\,d\theta.
\eeqn
In turn, (\ref{314}) is estimated by
$C\Vert Z^1\Vert^2_{L^2(\P)}$  by Proposition \ref{l4.1}, (ii). 
\hfill$\bo$

\begin{remark}\la{nonneg0}
The operator $\ti q_0(\theta)$ in $\HH^0$
is  nonnegative and self-adjoint.
Indeed, for any function $ Z\in {\cal D}$, we have
\beqn\nonumber
\int\limits_{K^d}
\Big(\ti q_0(\theta),\ti Z_\Pi(\theta,\cdot)\otimes
\overline{\ti Z_\Pi(\theta,\cdot)}\Big)
\,d\theta=
(2\pi)^{d}E|\langle Y,Z\rangle|^2\ge 0.
\nonumber
\eeqn
Hence, $\Big(\ti q_0(\theta),\ti Z_\Pi(\theta,\cdot)\otimes
\overline{\ti Z_\Pi(\theta,\cdot)}\Big)\ge 0$, $\theta\in K^d$.
\end{remark}

\setcounter{equation}{0}
\section{Compactness of measures $\mu_t$}

Proposition \re{l2.1}  follows
 from the bound (\ref{boundlocenergy})
by the Prokhorov Theorem \cite[Lemma II.3.1]{VF}
by using the method of \ci[Theorem XII.5.2]{VF}
because the embedding
${\cal E}^{0,\al}\subset {\cal E}^{s,\beta}$
is compact if $s<0$ and $\al>\beta$.
\begin{lemma}
Let conditions {\bf S0}--{\bf S3} hold.
Then the bounds (\ref{boundlocenergy})  hold
for $\al<-d/2$.
\end{lemma}
{\bf Proof.}
{\it Step (i).}
By  condition {\bf S1},
\beqn\la{4.2'}
E\Vert Y(t)\Vert^2_{0,\al}
&=&E\Big[
\int\limits_{\R^d}(1+|x|^2)^\al\Big(
|\psi(x,t)|^2+|\nabla\psi(x,t)|^2+|\pi(x,t)|^2\Big) \,dx
\nonumber\\
&&+ \sum_{k\in \Z^d}(1+|k|^2)^\al\Big(|u(k,t)|^2+|v(k,t)|^2\Big)\Big]\nonumber\\
&\le&C(\al,d) e(t),
\eeqn
where 
\beqn\nonumber
e(t):=E\Big[
\int\limits_{K_1^d}\Big(|\psi(y,t)|^2+|\nabla\psi(y,t)|^2+|\pi(y,t)|^2\Big) 
\,dy
+ |u(0,t)|^2+|v(0,t)|^2\Big].
\eeqn
Denote by $C_t$ the 
 correlation operator  of the random function
$$
\Big(\psi(y,t)|_{y\in K_1^d}, u(0,t), \pi(y,t)|_{y\in K_1^d}, v(0,t)\Big)
\in {\cal E}_1:=H^1(K_1^d)\oplus
 \R^d\oplus H^0(K_1^d)\oplus\R^d.
$$
Then  $e(t)$ is equal to the trace of the 
operator $C_t$.
Note that  $C_t={\rm Op}\Big(q_t(0,r,r')\Big)$
is an integral operator with the integral kernel
$ q_t(0,r,r')$ (see (\ref{qt})),
\be\la{4.1'}
q_t(0,r,r')=(2\pi)^{-d}\int\limits_{K^d}
e^{-i(r-r')\theta}\ti q_t(\theta,r,r')\,d\theta,\,\,\,\,
r,r'\in {\cal R}_1:=K_1^d\cup\{0\}.
\ee
Denote by $\ti q_t(\theta):={\rm Op}\Big(\ti q_t(\theta,r,r')\Big)$ 
the integral operator with the integral kernel $\ti q_t(\theta,r,r')$.
In this case, (\ref{4.1'}) implies that
\be\la{4.11}
e(t)=\tr_{{\cal E}_1} C_t=
(2\pi)^{-d}\int\limits_{K^d}
\tr_{{\cal E}_1}\Big[
e^{-ir\theta}\ti q_t(\theta)e^{ir'\theta}\Big]\,d\theta.
\ee
{\it Step (ii).}
Introduce the   operator $\Gamma$ defined by
$
\Gamma\ti\psi_\Pi(\theta,y):=
(\nabla_y\ti\psi_\Pi(\theta,y),\ti\psi_\Pi(\theta,y))
$
and  the operator $\Gamma_{ex}$ given by
\beqn
\nonumber
\Gamma_{ex}:=\left(\ba{cc}
\left(\ba{ll}\Gamma&0\\
0&1\ea\right)
&0\\
0&I\ea\right).
\eeqn
Then, since $\ti q_t(\theta)\ge 0$ and the operator
$e^{ir\theta}:(\psi(y),u,\pi(y),v)\to (e^{iy\theta}\psi,u,e^{iy\theta}\pi,v)$
in ${\cal E}_1$ is bounded  uniformly
with respect to $\theta\in K^d$ , we have
\be\la{4.12}
\tr_{{\cal E}_1}
\Big[e^{-ir\theta}\ti q_t(\theta)e^{ir'\theta}\Big]
\le
C\tr_{{\cal E}_1}\ti q_t(\theta)=C
\tr_{\HH^0}\Big[\Gamma_{ex}\ti q_t(\theta)\Gamma_{ex}^*\Big].
\ee
Let us now estimate the trace
 $\tr_{\HH^0}\Big[\Gamma_{ex}\ti q_t(\theta)\Gamma_{ex}^*\Big]$.
Introduce the matrix-valued self-adjoint
operator  $\Om_{ex}$ on the  space $\HH^0$,
$$
\Om_{ex}\equiv\Om_{ex}(\theta):=
\left(\ba{ll}
\Om(\theta)&0\\
0&I\ea\right),
$$
where $I$ stands for the identity operator on $H_1^0$.
Note that  $\Om_{ex}\ti q_t(\theta)\Om_{ex}\ge0$
(recall that $\Om_{ex}$ is a self-adjoint operator).
Further,
$B:=(\Gamma_{ex}\Om^{-1}_{ex})$ is a 
bounded operator on $\HH^0$ 
since
$\Om^{-1}: H^0(K_1^d)\oplus \C^n=
H^0_1\to H^1_1$,
and
$\left(\ba{ll}\Gamma&0\\
0&1\ea
\right): H_1^1\to H^0_1$.
Therefore,
\beqn\la{LambdaqLambda}
\tr_{\HH^0}[\Gamma_{ex}\ti q_t(\theta)\Gamma^*_{ex}]=
\tr_{\HH^0}[B\Om_{ex}\ti q_t(\theta)\Om_{ex}B^*]
\le
 C\tr_{\HH^0}[\Om_{ex}\ti q_t(\theta)\Om_{ex}],
\eeqn
by \ci[Theorem 1.6]{S}. Let us now
estimate the trace of the operator $\Om_{ex}\ti q_t(\theta)\Om_{ex}$.
We first use the formula $\Om_{ex}G(\theta,t)=U(\theta,t)\Om_{ex}$,
where $G(\theta,t)$ is defined in (\ref{G(t,theta)}) and
$$
U(\theta,t):= \left(\ba{rl}
\cos \Om t&\sin \Om t\\
-\sin\Om t&\cos \Om t
\ea \right).
$$
Hence, by  (\ref{ap1'}) we have
$$
\Om_{ex}\ti q_t(\theta)\Om_{ex}=
\Om_{ex}G(\theta,t)\ti q_0(\theta)G^*(\theta,t)\Om_{ex}
=U(\theta,t)\Om_{ex}\ti q_0(\theta)\Om_{ex}U^*(\theta,t).
$$
Since $U(t,\theta)$ is a unitary operator on $\HH^0$,
\be\la{trace}
\tr_{\HH^0}[\Om_{ex}\ti q_t(\theta)\Om_{ex}]=
\tr_{\HH^0}[U(\theta,t)\Om_{ex}\ti q_0(\theta)\Om_{ex}U^*(\theta,t)]=
\tr_{\HH^0}[\Om_{ex}\ti q_0(\theta)\Om_{ex}]
\ee
by \ci[Theorem VI.18, (c)]{RS1}.
Finally, it is follows from (\ref{4.11})-(\ref{trace}) that
\be\la{4.13}
\sup_{t\in\R}e(t)\le C_1
\int\limits_{K^d}\tr_{\HH^0}[\Om_{ex}\ti q_0(\theta)\Om_{ex}]\,d\theta.
\ee
{\it Step (iii).}
Let us now  prove that the RHS of (\ref{4.13}) is finite.
We use the representation  
$$
\Om_{ex}\ti q_0(\theta)\Om^*_{ex}=
(\Om_{ex}\Gamma^{-1}_{ex})\Gamma_{ex}\ti q_0(\theta)\Gamma^*_{ex}(\Om_{ex}\Gamma^{-1}_{ex})^*,
$$
where $\Gamma^{-1}_{ex}$
stands for the left inverse operator of $\Gamma_{ex}$.
On the other hand, $\Om_{ex}\Gamma^{-1}_{ex}$
is a bounded operator in $\HH^0$ since 
$\Om_{ex}(\theta)$ ($\Gamma^{-1}_{ex}$, resp.)
is (a finite - dimensional perturbation of) a 
 pseudodifferential operator  of order $1$ ($-1$, resp.) 
on $K_1^d$. Moreover, $\Om_{ex}(\theta)$
is uniformly bounded in $\theta\in K^d$.
Hence, 
\be\la{4.14}
\tr_{\HH^0}[\Om_{ex}\ti q_0(\theta)\Om_{ex}]\le
C\tr_{\HH^0}[\Gamma_{ex}\ti q_0(\theta)\Gamma^*_{ex}]=
C\tr_{{\cal E}_1}[\ti q_0(\theta)].
\ee
Finally, by inequalities (\ref{4.13}) and (\ref{4.14})
and by condition {\bf S2}, we obtain
\beqn\la{4.15}
\sup_{t\in\R}e(t)&\le& 
C\int\limits_{K^d}\tr_{{\cal E}_1}\ti q_0(\theta)\,d\theta\nonumber\\
&\le&C_1E\Big[
\int\limits_{K_1^d}(|\psi_0(y)|^2+|\nabla\psi_0(y)|^2+|\pi_0(y)|^2) \,dy
+ |u_0(0)|^2+|v_0(0)|^2\Big]
\nonumber\\
&\le& C_1(\bar e_F+e_L)<\infty.
\eeqn
Now the bound (\ref{boundlocenergy}) 
follows from (\ref{4.2'}) and (\ref{4.15}).
\hfill$\bo$

\setcounter{equation}{0}
\section{``Cutting out'' the  critical spectrum }
\begin{definition}\la{dC}
i) Introduce the critical set ${\cal C}:={\cal C}_*\cup
\Big(\cup_k{\cal C}_k\Big)$ (see {\bf E1}).\\
ii)
Introduce the set ${\cal D}^0\subset {\cal D}$ given by
\be\la{D0}
{\cal D}^0=\cup_{N}{\cal D}_{N},\,\,\,
{\cal D}_{N}:=\left\{Z\in{\cal D}\,
\left|
\ba{l}
P_l \ti Z_\Pi(\theta,\cdot)=0\,\,\,\mbox{for }\forall l\ge N,\,\theta\in K^d,\\
 \ti Z_\Pi(\theta,\cdot)=0\,\, \mbox{in a neighborhood of a set }\,{\cal C}
\cup \partial K^d.
\ea\right.\right\}
\ee
\end{definition}
\begin{lemma}\la{lcc}
Let $\lim\limits_{t\to\infty}{\cal Q}_t(Z,Z)={\cal Q}_\infty(Z,Z)$ 
for any $Z\in {\cal D}^0$.
Then the convergence holds for any $Z\in {\cal D}$.
\end{lemma}
{\bf Proof.}
First,  Definition \ref{dcorf} implies that
\be\la{Qtpsipsi}
{\cal Q}_t(Z,Z):=
E|\langle Y(\cdot,t),Z\rangle|^2=
\langle Q_t(p,p'),Z(p)\otimes Z(p')\rangle, \,\,\,Z\in {\cal D}.
\ee
Therefore, by (\ref{defY}) we have
${\cal Q}_t(Z,Z)={\cal Q}_0(Z(\cdot,t),Z(\cdot,t))$, and hence
\be\la{supQ}
\sup\limits_{t\in\R}|{\cal Q}_t(Z,Z)|\le
C\sup\limits_{t\in\R}\Vert Z(\cdot,t)\Vert^2_{{\bf L}^2}
\ee
by Corollary   \ref{c4.1}.
By the Parseval identity and  by the bound
(\ref{9.20'}), we obtain
\beqn\la{Phi(t)}
\Vert Z(\cdot,t)\Vert^2_{{\bf L}^2}&=&C(d)
\int\limits_{K^d}\Vert e^{\ti{\cal A}^T(\theta)t}\ti Z_\Pi(\theta,\cdot)
\Vert^2_{H_1^0\oplus H^0_1}\,d\theta\le C\int\limits_{K^d}
\Vert  \ti Z_\Pi(\theta,\cdot)\Vert^2_{H_1^0\oplus H_1^1}
\,d\theta \nonumber\\
&=&C \Vert Z\Vert^2_{{\cal L}} 
\eeqn
uniformly with respect to $t$. Here
${{\cal L}} :=
L^2(\R^d)\oplus[l^2(\Z^d)]^n\oplus
H^1(\R^d)\oplus[l^2(\Z^d)]^n$.
Further, by Lemma \ref{lc*},
for any $Z\in{\cal D}$, we can find a 
$Z^{N}\in{\cal D}_{N}$ such that
$\Vert Z-Z^{N}\Vert_{{\cal L}}\to 0$ as $N\to\infty$. 
Namely, $\ti Z^N_\Pi(\theta)
=\sum_{l\le N}P_l(\theta)\ti Z_\Pi(\theta)$
if $\ti Z_\Pi(\theta)=0$ for $\theta$ in a
neighborhood  of ${\cal C}\cup \partial K^d $.
Finally, the set of such functions $Z$
is dense in ${\cal L}$.
Then Lemma \ref{lcc} follows from
(\ref{supQ}), (\ref{Phi(t)}), and Corollary \ref{c4.2}.
\hfill$\bo$

\begin{lemma}\la{lck}
The convergence (\ref{2.6i}) holds
for any $Z\in{\cal D}$
if it holds for $Z\in{\cal D}^0$.
\end{lemma}
{\bf Proof.} This  follows immediately from
 Lemma \re{lcc} by the Cauchy-Schwartz inequality:
$$
\ba{rcl}
|\hat\mu_t(Z')-\hat\mu_t(Z'')|&=&
|\ds\int \Big( e^{i\langle Y,Z' \rangle}-
e^{i\langle Y,Z'' \rangle}\Big)\mu_t(dY)|
\le
\ds\int |e^{i\langle Y,Z'-Z'' \rangle}-1|\mu_t(dY)\\
~\\
&\le&
\ds\int |\langle Y,Z'-Z'' \rangle|\mu_t(dY)
\le \sqrt {\ds\int |\langle Y,Z'-Z'' \rangle|^2
\mu_t(dY)}\\
~\\
&=&
\sqrt {{\cal Q}_t(Z'-Z'', Z'-Z'')}
\le C\Vert Z'-Z'' \Vert_{{\cal L}}.
~~~~~~~~~~\bo
\ea
$$

\setcounter{equation}{0}
\section{ Convergence of the covariance}
\begin{pro}\la{p4.1}
Let conditions {\bf E1}-{\bf E2}, {\bf R1}-{\bf R3}
  and {\bf S0}-{\bf S3} hold.
Then, for any  $Z\in {\cal D}$,
\be\la{4.4}
{\cal Q}_t(Z,Z)
\to {\cal Q}_\infty(Z,Z),\,\,\,\,t\to \infty.
\ee
\end{pro}
{\bf Proof.}  By Lemma~\ref{lcc}, 
it suffices to prove the convergence (\ref{4.4})
 for $Z\in{\cal D}^0$ only.
If $Z\in {\cal D}^0$, then  $Z\in {\cal D}_{N}$
for some $N$.
Let us  apply the Zak transform to the matrix
$Q_{t}(p,p')$,
\be\la{6.200}
\langle Q_t(p,p'),Z(p)\otimes Z(p')\rangle
=(2\pi)^{-2d}
\langle \ti Q_t(\theta,\theta',r,r'),
\ti Z_\Pi(\theta,r)
\otimes\ov{\ti Z}_\Pi(\theta',r')\rangle.
\ee
Further, by Lemma \ref{lc*},
we can choose some
smooth branches of the functions $F_l(\theta,r)$ and $\om_l(\theta)$
to apply the stationary phase arguments, which
requires some smoothness with respect to $\theta$.
Denote by $\supp \ti Z_\Pi$ the closure
of the set 
$\{\theta\in K^d: \ti Z_\Pi(\theta,y)\not\equiv 0,y\in T_1^d\}$.
Since $\supp \ti Z_\Pi\cap ({\cal C} \cup \partial K^d)=\emptyset$,
we can apply Lemma \ref{lc*}. Namely, for any point 
$\Theta\in \supp \ti Z_\Pi$, there is a neighborhood
${\cal O}(\Theta)\subset K^d\setminus({\cal C} \cup \partial K^d)
$ with the corresponding properties.
Hence,
$\supp \ti Z_\Pi\subset \cup_{m=1}^M {\cal O}(\Theta_m)$,
where $\Theta_m\in \supp \ti Z_\Pi$.
Therefore, there is a  finite partition of unity 
\be\la{part}
\sum_{m=1}^M g_m(\theta)=1,\,\,\,\,\theta\in \supp
\ti Z_\Pi,
\ee
where $g_m$ are nonnegative functions of 
$C_0^\infty(K^d)$ and
$\supp g_m\subset {\cal O}(\Theta_m)$.
Further, using Definition \ref{dC}, ii) and the
partition (\ref{part}), represent
the  RHS of (\ref{6.200}) as
\beqn\la{6.20}
\langle Q_t(p,p'),Z(p)\otimes Z(p')\rangle
=(2\pi)^{-d}\sum\limits_{m=1}^{M}\sum\limits_{l,l'=1}^{N}
\langle g_m(\theta)r_{ll'}(t,\theta), A_l(\theta)
\otimes\ov{A}_{l'}(\theta)\rangle,
\eeqn
 using formulas (\ref{ap22}) and (\ref{6.10}).
Here
$A_l(\theta)=(F_l(\theta,\cdot), \ti Z_\Pi(\theta,\cdot))$, 
$r_{ll'}(t,\theta)$ is
the $2\times 2$ matrix 
\beqn\la{ap.2}
r_{ll'}(t,\theta)&:=&\frac{1}{2}\sum\limits_{\pm}\Big\{
\cos\big(
\om_l(\theta)\!\pm\!\om_{l'}(\theta)\big)t
~\Big(p_{ll'}(\theta)
\mp C_l(\theta)p_{ll'}(\theta)
C^T_{l'}(\theta)\Big)
\nonumber\\
&&+
\sin\big(\om_l(\theta)\!\pm\!\om_{l'}(\theta)\big)t
~\Big(
 C_l(\theta)p_{ll'}(\theta)\pm p_{ll'}(\theta)
C_{l'}^T(\theta)\Big)\Big\},
\eeqn
where
\beqn\la{C(theta)}
& C_l(\theta):=\left(
\ba{lcl}
0&\om^{-1}_l(\theta)\\
-\om_l(\theta)&0
\ea
\right),\,\,\,\,
 C^T_l(\theta):=\left(
\ba{lcl}
0&-\om_l(\theta)\\
\om^{-1}_l(\theta)&0
\ea
\right),\\
\la{pij}
&p_{ll'}^{ij}(\theta):=
\Big(F_l(\theta,\cdot),
( \ti q_0^{ij}(\theta)F_{l'})(\theta,\cdot)\Big),\,\,\,\,
\theta\in {\cal O}(\Theta_m),
\,\,\,\, l,l'=1,2,\dots,\,\,\,\, i,j=0,1,
\eeqn
and $(\cdot,\cdot)$ stands for the inner product in
$H^0_1\equiv H^0(T_1^d)\oplus \C^n$ (see (\ref{H_1^0}))
or in $\HH^0\equiv[H^0_1]^2$.
By Lemma \re{lc*}, 
 the eigenvalues $\om_l(\theta)$ 
and the eigenfunctions $F_l(\theta,r)$  
are  real-analytic functions in 
 $\theta\in\supp g_m$ for every $m$: we do not mark the 
functions by the index $m$ to simplify the notation.
\begin{lemma}\la{estpll''}
Let conditions {\bf S0}--{\bf S3} hold.
Then  $p^{ij}_{ll'}(\theta)\in L^1({\cal O}(\Theta_m))$,  $i,j=0,1$,
$l,l'=1,2,\dots$ for each $m=1,\dots, M$.
\end{lemma}
{\bf Proof}.
Since $\{F_l(\theta,\cdot)\}$ is an orthonormal basis,
 by the Cauchy-Schwartz inequality, we have
\beqn
\Big|\int\limits_{{\cal O}(\Theta_m)}
|p_{ll'}^{ij}(\theta)|\,d\theta\Big|^2&\le&
C\int\limits_{{\cal O}(\Theta_m)}
|p_{ll'}^{ij}(\theta)|^2\,d\theta\le
\int\limits_{{\cal O}(\Theta_m)}\Big|\Big(F_l(\theta,r),
\ti q_0^{ij}(\theta,r,r')F_{l'}(\theta,r')\Big)\Big|^2\,d\theta
\nonumber\\
&\le&
\int\limits_{{\cal O}(\Theta_m)}
d\theta\int\limits_{{\cal R}}dr\int\limits_{{\cal R}}
|\ti q^{ij}_0(\theta,r,r')|^2\,dr'.\nonumber\bo
\eeqn
Further, let us study the terms in (\ref{6.20}),
which are oscillatory integrals with respect to the variable
 $\theta$.
 The identities
$\om_l(\theta)+\om_{l'}(\theta)\equiv\const_+$
or $\om_l(\theta)-\om_{l'}(\theta)\equiv\const_-$
 with  $\const_\pm\ne 0$
are impossible by   condition {\bf E2}.
Moreover, the oscillatory integrals
with $\om_l(\theta)\pm \om_{l'}(\theta)
\not\equiv \const$
vanish as $t\to\infty$. Hence,  only the integrals with
$\om_l(\theta)-\om_{l'}(\theta)\equiv 0$
contribute to the limit because the relation
  $\om_l(\theta)+\om_{l'}(\theta)\equiv 0$
 would imply the relation
$\om_l(\theta)\equiv\om_{l'}(\theta)\equiv 0$, which
is impossible by  {\bf E2}.
Let us index the eigenvalues $\om_l(\theta)$
as in (\ref{enum}). Then
$\cos\big(\om_l(\theta)-\om_{l'}(\theta)\big)t=1$
 for  $l,l'\in(r_{\sigma-1},r_\sigma]$,
 $\sigma=1,2,\dots$. Hence,  for  $l,l'\in(r_{\sigma-1},r_\sigma]$, we have
\beqn\la{ap.3}
r_{ll'}(t,\theta)&=&
\frac{1}{2}\Big(
p_{ll'}(\theta)+ C_l(\theta)p_{ll'}(\theta)C_{l'}^T(\theta)\Big)
\nonumber\\
&&+ \frac{1}{2}\cos 2\om_l(\theta)t
~\Big(p_{ll'}(\theta)- C_l(\theta)p_{ll'}(\theta)C_{l'}^T(\theta) \Big)
\nonumber\\
&&+ \frac{1}{2}\sin2\om_l(\theta)t
~\Big( C_l(\theta)p_{ll'}(\theta)\!+\!
p_{ll'}(\theta)C_{l'}^T(\theta)\Big).
\eeqn
Therefore,
\beqn\la{3.10}
\langle Q_t(p,p'),Z(p)\otimes Z(p')\rangle
=(2\pi)^{-d}\sum_m\sum\limits_{l,l'=1}^{N}\int
g_m(\theta)\Bigl( M_{ll'}(\theta)
+\dots,A_l(\theta)\otimes\ov{A}_{l'}(\theta)\Big)\,d\theta,
\eeqn
where
$M_{ll'}(\theta)=(M^{ij}_{ll'}(\theta))_{i,j=0}^{1}$,
$l,l'=1,2,\dots,$ is the  matrix
with the continuous  entries
\beqn\la{Piinfty}
M^{ij}_{ll'}(\theta)=\chi_{ll'}\frac{1}{2}
\Big(F_l(\theta,r),
\Big[\ti q_0(\theta,r,r')+
 C_l(\theta)\ti q_0(\theta,r,r') C^T_{l}(\theta)\Big]^{ij}
F_{l'}(\theta,r')\Big);
\eeqn
here the symbol $\chi_{ll'}$ is given by
(see (\re{enum}))
\be\la{chi}
\chi_{ll'}:=\left\{\ba{rl}
1 &{\rm if} \,\,\,\, l,l'\in(r_{\si-1}, r_{\si}],\,\,\,
\si=1,2,...,\,\,r_0:=0, \\
 0& {\rm otherwise}.
\ea\right.
\ee
Further, for  $\theta\in \supp g_m\subset {\cal O}(\Theta)$
  (see Lemma \ref{lc*}), we write 
\be\la{qinfty}
\ti q_\infty^{ij}(\theta,r,r')=
\sum\limits_{l,l'=1}^{+\infty}F_l(\theta,r)
{M_{ll'}^{ij}}(\theta)\overline{F_{l'}}
(\theta,r'),
\,\,\,\,i,j= 0,1.
\ee
The local representation (\re{qinfty}) can be expressed globally in the form (\ref{qinfty+}). Hence,
\beqn\la{3.10'}
\langle Q_t(p,p'),Z(p)\otimes Z(p')\rangle
= (2\pi)^{-d}\sum_m\int
g_m(\theta)\Bigl(\ti q_{\infty}(\theta,r,r'),
\ti Z_\Pi(\theta,r)\otimes\ov{\ti Z}_\Pi(\theta,r')\Big)\,d\theta\! +\!\dots,
\eeqn
where the symbol  $"\dots"$ stands for the oscillatory
 integrals which contain
 $\cos(\om_l(\theta)\pm\om_{l'}(\theta))t$
and $\sin(\om_l(\theta)\pm\om_{l'}(\theta))t$
with $\om_l(\theta)\pm\om_{l'}(\theta)\not\equiv$const.
The oscillatory integrals converge to zero  by the
Lebesgue-Riemann Theorem because the integrands in ``$...$''
are summable, and we have
$\na(\om_l(\theta)\pm\om_{l'}(\theta))=0$ 
 on the set of Lebesgue measure zero only.
The summability follows from Lemma \ref{estpll''} because
 the functions $A_l(\theta)$ are smooth.
The zero-measure condition follows as in (\re{c*})
since $\om_l(\theta)\pm\om_{l'}(\theta)\not\equiv$const.
This completes the proof of  Proposition \ref {p4.1}.
\hfill$\bo$

\setcounter{equation}{0}
\section{Bernstein's argument}
 \subsection{Oscillatory
 representation and stationary phase method}

To prove  (\ref{2.6i}), we evaluate $\langle Y(\cdot,t),Z\rangle$
by duality arguments. Namely, 
introduce the dual space ${\cal E}':=H^{-1,-\al}(\R^d)\oplus
L^{-\al}\oplus H^{0,-\al}(\R^d)\oplus L^{-\al}$
 with  finite  norm
\beqn\nonumber
(\bre Z\bre'_{0,-\al})^2:=
\Vert \psi\Vert_{-1,-\al}^2+
\Vert\pi\Vert_{0,-\al}^2+\Vert u\Vert_{-\al}^2+
\Vert v\Vert_{-\al}^2.
\eeqn
For  $t\in\R$, introduce the ``formal adjoint'' 
operator $W'(t)$,
\be\la{defW}
\langle W(t)Y,Z\rangle:=
\langle Y,W'(t)Z\rangle,\,\,\,Y\in{\cal E},
\,\,\,Z\in{\cal E}',
\ee
where $\langle\cdot,\cdot\rangle$
stands for the inner product in
$ L^2(\R^d)\oplus [l^2(\Z^d)]^n\oplus L^2(\R^d)\oplus [l^2(\Z^d)]^n$.

 Write $Z(\cdot,t)=W'(t)Z$. Then
formula (\ref{defW}) can be rewritten as
\be\la{defY}
\langle Y(t),Z\rangle =\langle Y_0,Z(\cdot,t)\rangle,
\,\,\,\,t\in\R.
\ee
The adjoint group $W'(t)$
admits a convenient description.
\begin{lemma}\la{ldu}
The action of the group $W'(t)$ coincides  with the action
of  $W(t)$ up to the order of  components.
Namely, $W'(t)=\exp({\cal A}^Tt)$,
where ${\cal A}$ is the generator of the group $W(t)$.
\end{lemma}
{\bf Proof.}
Differentiating (\ref{defW}) with respect to $t$ for
$Y,Z\in {\cal D}$, we obtain
\be\la{UY}
\langle Y,\dot W'(t)Z\rangle =\langle
\dot W(t)Y,Z\rangle .
\ee
The group $W(t)$ has the generator ${\cal A}$
(see (\ref{A})).
The generator of $W'(t)$   is the conjugate operator
\be\la{A'}
{\cal A}'=\left(\ba{cc}
0 &-{\cal H}'\\
 1 & 0\ea\right),
\,\,\,\,{\cal H}'={\cal H}=
\left(\ba{cc}
-\De+m_0^2   & S\\
S^*  &-\De_L+\nu_0^2 \ea
\right).
\ee
Hence, ${\cal A}'={\cal A}^T$.
\hfill$\bo$
\begin{cor}
The following uniform bound holds:
\be\la{9.20'}
\Vert e^{\ti{\cal A}^T(\theta)t}
\ti Z_\Pi(\theta,\cdot)\Vert_{H^0_1\oplus H_1^1}
\le C\Vert \ti Z_\Pi(\theta,\cdot)\Vert_{H^0_1\oplus H_1^1},
\,\,\,\ti Z_\Pi(\theta,\cdot)\in H^0_1\oplus H_1^1,
\ee
which can be proved  similarly to (\ref{10.11}).
\end{cor}

Applying Lemma \ref{ldu},
we can rewrite $Z(t)=W'(t)Z$ as the Zak transform, i.e.,
$\ti Z_\Pi(\theta,r,t)=
\exp\Big(\ti{\cal A}^T(\theta) t\Big)\ti Z_\Pi(\theta,r).$
Recall that we can restrict ourselves to elements
$Z\in{\cal D}_{N}$ with a fixed index $N$.
Using the partition of unity (\ref{part}), we obtain 
\beqn\la{frepe}
Z(k+r,t)&=&(2\pi)^{-d}
\sum\limits_{m=1}^M\sum\limits_{l=1}^{N}
\int\limits_{K^d} g_m(\theta)
e^{-i(k+r)\theta}  G^T_l(\theta,t)
F_l(\theta,r) A_l(\theta)\,d\theta\nonumber\\
&=&\sum_{m,\pm}\sum\limits_{l=1}^{N}
\int\limits_{K^d}
e^{-i(\theta (k+r)\pm\om_l(\theta) t)}
g_m(\theta)a^\pm_l(\theta)
F_l(\theta,r)A_l(\theta)\,d\theta,
\,\,\,\,Z\in{\cal D}_N.
\eeqn
Here
$A_l(\theta)=(F_l(\theta,\cdot),
\ti Z_{\Pi}(\theta,\cdot))$,
\beqn \la{G_l}
 G_l(\theta,t)&:=&\left(
\ba{cc}
\cos\om_l(\theta)t&\ds\frac{\sin\om_l(\theta)t}{\om_l(\theta)}\\
-\om_l(\theta)\sin\om_l(\theta)t&\cos\om_l(\theta)t
\ea\right),\,\,\,\,
\theta\in \supp g_m,
\eeqn
and $\om_l(\theta)$ and $a^\pm_l(\theta)$
are real-analytic functions in the interior of the set
 $\supp g_m$ for every $m$.

Let us  derive formula (\ref{2.6i}) by analyzing the
propagation of the solution $Z(k+r,t)$
of the form (\ref{frepe}) in diverse
directions $k=vt$ with $v\in\R^d$ and for $r\in{\cal R}$.
To this end, we apply the stationary phase method
 to the oscillatory integral (\ref{frepe})
along the rays $k=vt$, $t>0$. Then the phase  becomes
$(\theta v\pm\om_l(\theta))t$, and its stationary
 points are the solutions of
the equations $v=\mp\nabla\om_l(\theta)$.

Note that $\ti Z_\Pi(\theta,r)=0$ at
 the points $(\theta,r)\in K^d\oplus {\cal R}$
with  degenerate Hessian $D_l(\theta)$ (see {\bf E1}).
Therefore, the stationary phase method leads to the following
 two different types  of 
asymptotic behavior of $Z(vt,t)$ as $t\to\infty$.
\smallskip\\
{\bf I.} Let  the velocity $v$ be inside the light cone,
$v=\pm\nabla\om_l(\theta)$, where
$\theta\in {\cal O}(\Theta)\setminus {\cal C}$.
Then
\be\la{spd}
 Z(vt,t)={\cal O}( t^{-d/2}).
\ee
{\bf II.} Let  the velocity $v$ be outside the light cone,
$v\ne\pm\na\om_l(\theta)$, where
$\theta\in  {\cal O}(\Theta)\setminus{\cal C}$, $l=1,\dots,N$.
Then
\be\label{rd}
Z(vt,t)={\cal O}( t^{-k}),\,\,\,\, \forall k>0.
\ee
\begin{lemma}\la{l5.3}
The following
bounds hold for any fixed $Z \in {\cal D}^0$:
\be\la{bphi}
\!\!\! i)~~~~~
\sup_{p\in \P}|Z(p,t)| \le  C~t^{-d/2}.
~~~~~~~~~~~~~~~~~~~~~~~~~~~~~~~~~~~~~~~~~~~~~~~~~~~~~~~~
\ee
ii) For  any $k>0$, there exist
numbers $C_k,\ga>0$ such that
\be\la{conp}
|Z(p,t)|\le C_k(1+|p|+|t|)^{-k},\quad\quad
 |p|\ge\gamma t.
\ee
\end{lemma}
{\bf Proof.}
Consider $Z(k+r,t)$ along each ray
$k=vt$ with  an arbitrary $v\in\R^d$.
Substituting  the related expressions into 
 (\ref{frepe}), we obtain
\be\la{freper}
Z(vt+r,t)=\sum_{m,\pm} \sum\limits_{l=1}^{N}
\int\limits_{K^d}
e^{-i(\theta v\pm\om_l(\theta)) t}
e^{-i\theta r}a^\pm_l(\theta)
F_l(\theta,r)A_l(\theta)\,
d \theta,\,\,\,\, Z\in{\cal D}_{N}.
\ee
This is a sum of oscillatory integrals with  phase
functions of the form $\phi_l^\pm(\theta)=
\theta v\pm\om_l(\theta)$
and with amplitudes $a^\pm_l(\theta) $
that are real-analytic functions of  $\theta$
in the interiors of the sets $\supp g_m$.
Since $\om_l(\theta)$ is real-analytic,
 each function $\phi_l^\pm$  has at most
 finitely many  stationary points $\theta\in\supp g_m$
(solutions of the equation $v=\mp\nabla\om_l(\theta)$).
The stationary points are nondegenerate for
$\theta\in\supp g_m$
by  Definition \ref{dC} and by {\bf E1}
since
\be\la{Hess}
{\rm det}\Big(\frac{\pa^2 \phi_l^\pm}
{\pa \theta_i\pa \theta_j}\Big)=
\pm D_l(\theta)\not= 0,\,\,\,\,\,\theta\in \supp g_m.
\ee
At last, $\ti Z_\Pi(\theta,r)$ is smooth because
  $Z\in{\cal D}$.
Therefore, we have $Z(vt+r,t)={\cal O}(t^{-d/2})$
 according to the standard stationary phase method
  of \ci{F, RS3}.  This implies the bounds (\ref{bphi})
in each cone $|k|\le ct$ with any finite $c$.

 Further, write
$\bar v:=\max_m\max_{l=1,N}
\max\limits_{\theta\in \supp g_m}
|\nabla \om_l(\theta)|.$
Then, for $|v|>\bar v$, there are no
 stationary points  in  $\supp \ti Z_\Pi $.
Hence,  integration by parts (as in \ci{RS3}) yields
$Z(vt+r,t)={\cal O}(t^{-k})$ for any $k>0$.
On the other hand, the integration by parts in
(\ref{frepe}) implies a  similar bound,
$Z(p,t)={\cal O}\Big(\ds(t/|p|)^l\Big)$ for any $l>0$.
Therefore, relation (\ref{conp}) follows with any
$\ga>\ov v$.
This shows that the bounds (\ref{bphi}) hold everywhere.
\hfill$\bo$

\subsection{``Room-corridor'' partition}
The remaining constructions in the proof of (\ref{2.6i}) are
 similar to \ci{DKKS, DKS1}.
However, the proofs are not identical, since here we consider
a non-translation-invariant case and a coupled system.

Introduce a ``room-corridor''  partition of the
ball $\{p\in \P:~|p|\le \ga t\}$  with $\ga$ taken
from (\ref{conp}).
For $t>0$,  choose $\De_t$ and $\rho_t\in\N$.
Asymptotic relations between $t$, $\De_t$ and  $\rho_t$
are specified below.
Set $h_t=\De_t+\rho_t$ and
\be\la{rom}
a^j=jh_t,\,\,\,b^j=a^j+\De_t,\,\,\,
j\in\Z,\,\,\,\,\,\,N_t=[(\gamma t)/h_t].
\ee
The slabs $R_t^j=\{p\in \P:
 |p|\le N_t h_t,\,a^j\le p_d< b^j\}$
are referred to as  ``rooms'',
$C_t^j=\{p\in \P: |p|\le
N_t h_t,\, b^j\le p_d<  a^{j+1}\}$  as ``corridors'',
and $L_t=\{p\in \P: |p|> N_t h_t\}$ as ``tails''.
Here  $p=(p_1,\dots,p_d)$,
$\De_t$ is the width of a room, and
$\rho_t$  is that of a corridor.
Denote  by  $\chi_t^j$ the indicator of the room $R_t^j$,
by $\xi_t^j$ the indicator of the corridor $C_t^j$, and
by $\eta_t$ the indicator of  the tail $L_t$. In this case,
\be\la{partB}
{\sum}_t
[\chi_t^j(p)+\xi_t^j(p)]+ \eta_t(p)=1,\,\,\,p\in \P,
\ee
where the sum ${\sum}_t$ stands for
$\sum\limits_{j=-N_t}^{N_t-1}$.
Hence, we obtain the following  Bernstein's type representation:
\be\la{res}
\langle Y_0,Z(\cdot,t)\rangle = {\sum}_t
[\langle Y_0,\chi_t^j Z(\cdot,t)\rangle +
\langle Y_0,\xi_t^jZ(\cdot,t)\rangle ]+
\langle Y_0,\eta_tZ(\cdot,t)\rangle.
\ee
Introduce  the random variables
 $ r_{t}^j$, $ c_{t}^j$ and $l_{t}$ by the formulas
\be\la{100}
r_{t}^j= \langle Y_0,\chi_t^j Z(\cdot,t)\rangle,~~
c_{t}^j= \langle Y_0,\xi_t^j Z(\cdot,t)\rangle,
\,\,\,l_{t}= \langle Y_0,\eta_t Z(\cdot,t)\rangle.
\ee
Then relation  (\ref{res}) becomes
\be\la{razli}
\langle Y_0,Z(\cdot,t)\rangle =
{\sum}_t (r_{t}^j+c_{t}^j)+l_{t}.
\ee

\begin{lemma}  \la{l5.1}
    Let {\bf S0--S3} hold and $Z\in{\cal  D}^0$.
The following bounds hold for $t>1$:
\beqn
E|r^j_{t}|^2&\le&  C(Z)~\De_t/ t,\,\,\,\forall j,\la{106}\\
E|c^j_{t}|^2&\le& C(Z)~\rho_t/ t,\,\,\,\forall j,\la{106''}\\
E|l_{t}|^2&\le& C_k(Z)~t^{-k},\,\,\,\,\forall k>0.
\la{106'''}
\eeqn
\end{lemma}
{\bf Proof.} Relation
 (\ref{106'''}) follows from (\ref{conp}) and
Proposition \ref{l4.1}, (i).
We discuss  (\ref{106}) only, and  relation (\ref{106''})
can be studed in a similar way.
Let us express $E|r_t^j|^2$  in terms of correlation matrices.
Definition (\ref{100})   implies
 \be\la{100rq}
E|r_{t}^j|^2= \langle Q_0(p,p'),
 \chi_t^j(p)Z(p,t)\otimes\chi_t^j(y)Z(p',t)\rangle.
\ee
 According to (\ref{bphi}),  Eqn (\ref{100rq}) yields
\beqn\la{er}
E|r_{t}^j|^2&\le&
C t^{-d}\int 
\chi_t^j(p)\Vert Q_0(p,p')\Vert\,dpdp' \nonumber\\
&=&Ct^{-d}\int \chi_t^j(p)\,dp
\int\Vert Q_0(p,p')\Vert\,dp'
\le C \De_t/t,
\eeqn
where $\Vert Q_0(p,p')\Vert $ stands for the norm of the matrix
$\left(Q_0^{ij}(p,p')\right)$.
Therefore,  (\ref{er})   follows by Corollary \ref{c4.1}.
\hfill$\bo$

\subsection{Proof of Theorem A}
The remaining part of the proof of  the convergence (\ref{2.6i}) uses  the Ibragimov-Linnik central limit theorem \ci{IL}
and the bounds (\ref{106})-(\ref{106'''}).
For details, see
 \ci[Sections 8,9]{DKKS} and \ci[Sections 9,10]{DKM1}.

\setcounter{equation}{0}
\section{ Ergodicity and mixing  for the limit
 measures}
The limit measure $\mu_\infty$ is invariant by Theorem~A, {\it(iv)}.
Let $E_\infty$ be the integral with respect to  $\mu_\infty$.
\begin{theorem}
Let the assumptions of Theorem A hold. Then
$W(t)$ is mixing with respect to
 the corresponding limit measure $\mu_\infty$, i.e.,
for any $f,g\in L^2({\cal E},\mu_\infty)$ we have
\be\la{3D}
\lim_{t\to\infty}
E_\infty f(W(t)Y)g(Y)=
 E_\infty f(Y)E_\infty g(Y).
\ee
In particular,
the group $W(t)$ is ergodic with respect to the measure $\mu_\infty$,
\be\la{4D}
\lim_{T\to\infty}
\frac{1}{T}\int\limits_0^T f(W(t)Y) dt=
E_\infty f(Y)~~(\mbox{mod }\mu_\infty).
\ee
\end{theorem}
{\bf Proof.} {\it Step (i).}
Since $\mu_\infty$ is Gaussian, the proof of (\ref{3D})
reduces to that of the convergence
\be\la{5D}
\lim_{t\to\infty}
E_\infty \langle W(t)Y,Z\rangle
\langle Y,Z^1\rangle=0.
\ee
for any $Z,Z^1\in {\cal D}$.
It suffices to prove relation (\ref{5D}) for $Z, Z^1\in{\cal D}_{N}$.
However, 
this  follows from Corollary \ref{c4.2} and Theorem A, {\it(iv)}.

{\it Step (ii).} Let $Z, Z^1\in{\cal D}_{N}$.
Applying  the Zak transform
and the Parseval identity, we obtain
\beqn\la{6D}
I(t)&\equiv&
E_\infty \langle W(t)Y,Z\rangle
\langle Y,Z^1\rangle =
E_\infty \langle Y,W'(t)Z\rangle
\langle Y,Z^1\rangle\nonumber\\
&=&(2\pi)^{-d}\langle
\ti q_\infty(\theta,r,r'),
G^*(\theta,t)\ti Z_\Pi(\theta,r)\otimes
\overline{\ti Z^1_\Pi}(\theta,r')\rangle.
\eeqn
Using  a finite partition of unity
(\ref{part}),
and relations (\ref{6D}) and (\ref{qinfty}), we see that
\be
I(t)=(2\pi)^{-d}\sum\limits_{m=1}^M
\sum\limits_{l,l'=1}^N
\int  g_m(\theta)G_l^*(\theta,t)A_l(\theta)M_{ll'}(\theta)
 \overline{A_{l'}(\theta)}\,d\theta.
\ee
Here $A_l(\theta)=(F_l(\theta,\cdot), \ti Z_\Pi(\theta,\cdot))$ and
$A_{l'}(\theta)=(F_l(\theta,\cdot), \ti Z^1_\Pi(\theta,\cdot))$, and
 $G_l(t,\theta)$ is defined in (\ref{G_l}).
Similarly to (\ref{frepe}), we  have
\be\la{sln}
I(t)=\sum_m \sum\limits_{l,l'=1}^N
\,\int
g_m(\theta)e^{\pm i\om_l(\theta )t}a_l^{\pm}(\theta)
A_l(\theta)M_{ll'}(\theta) \overline{A_{l'}(\theta)}\,d\theta.
\ee
Here all the phase functions $\om_l(\theta)$ and
the amplitudes $a_l^{\pm}(\theta)$
are smooth functions on  $\supp g_m$.
Further, the relation $\na\om_l(\theta)= 0$
holds  on a set of  Lebesgue measure zero only.
This follows similarly to (\re{c*})
since $\na\om_l(\theta)\not\equiv$const
by  condition {\bf E2}.
Hence, $I(t)\to 0$ as $t\to\infty$
by the Lebesgue-Riemann theorem since
the functions  $M_{ll'}(\theta)$ are continuous.
\hfill$\bo$\medskip\\
{\bf Remark.} A similar result for wave  equations  and for harmonic crystals
was  proved in \ci{DK, DKS1}.

\setcounter{equation}{0}
 \section{ Appendix A: Dynamics in the Bloch-Fourier representation}
In this appendix we prove the bound (\ref{boundsol}).
We first  construct the exponential $\exp(\ti{\cal A}(\theta)t)$
for any chosen $\theta\in K^d\equiv [0,2\pi]^d$ and study
 its properties.
Let us choose $\theta\in K^d$ and $X_{0}\in \HH^1:=H^1_1\oplus
H^0_1$, where $H^s_1\equiv H^s(T_1^d)\oplus \C^n$.
Introduce the functions $\exp(\ti{\cal A}(\theta)t)X_{0}$ for
$X_{0}\in \HH^1$ as the solutions $ X(\theta,t)$
to the problem 
\beqn\la{CPF'}
\left\{\ba{ll}
\dot X(\theta,t)=
\ti {\cal A}(\theta)X(\theta,t),\,\,\,t\in\R,\\
\,\,\,\,X(\theta,0)=X_{0}.
\ea\right.
\eeqn
\begin{pro}\la{c.a2}
For any chosen $\theta\in K^d$, the Cauchy problem (\ref{CPF'})
admits a unique solution $X(\theta,t)\in C(\R;\HH^1)$.
Moreover, 
\be\la{9.12'}
X(\theta,t)=e^{\ti{\cal A}(\theta) t}X_0,
\ee
and
\be\la{10.11}
\Vert X(\theta,t)\Vert_{\HH^1}\le
C\Vert X_0\Vert_{\HH^1},
\ee
where the constant $C$ does not depend on $\theta\in K^d$
and  $t\in\R$.
\end{pro}

We prove this proposition in Subsection 9.2.
\subsection{Schr\"odinger operator}
Let us first construct solutions
$X(\theta,t)$ to  problem (\ref{CPF'})
with a chosen parameter $\theta\in K^d$.
Write  $X(\theta,t)=(X^0(\theta,t),
X^1(\theta,t))$, where
$X^0(\theta,t)=(\varphi(\theta,t), u(\theta,t))$ and
$X^1(\theta,t)=(\phi(\theta,t),v(\theta,t))$.
By (\ref{CPF'}) and (\ref{tiA}) we have
$X^1(\theta,t)=\dot X^0(\theta,t)$,
and $X^0(\theta,t)$ is a solution to the following
Cauchy problem with a chosen parameter $\theta\in K^d$:
\beqn\la{Phit}
\left\{\ba{rcl}
\ddot X^0(\theta,t)&=&-\ti {\cal H}(\theta)X^0(\theta,t),\,\,\,\,t\in\R,\\
\Big(X^0(\theta,t),
\dot X^0(\theta,t)\Big)\Big|_{t=0}&=&
(X^0_{0},X^1_{0})=X_{0},
\ea\right.
\eeqn
where $\ti {\cal H}(\theta)$ is the ``Schr\"odinger operator''
(\ref{H(theta)}).  Hence, formally,
\be\la{rep}
X^0(\theta,t)=\cos\Om(\theta)t\, X^0_{0}+
\sin\Om(\theta)t\,\Om^{-1}(\theta)X^1_{0},
\ee
where $\Om(\theta)=\sqrt{\ti{\cal H}(\theta)}> 0$.
\begin{lemma}
For $X^0\in H_1^0$, 
the following bounds hold:
\beqn
\Vert \Om(\theta) X^0\Vert_{H_1^{-1}}&\le& C
\Vert  X^0\Vert_{H_1^{0}},\la{estOm}\\
\Vert \ti{\cal H}^{-1}(\theta) X^0\Vert_{H_1^{0}}&\le& C
\Vert X^0\Vert_{H_1^{0}},\la{estH-1}
\eeqn
where the  constant $C$ does not depend on $\theta\in K^d$.
\end{lemma}
{\bf Proof} (i)  Formula
(\ref{H(theta)}) for  $\ti {\cal H}(\theta)$ implies that
\be\la{10.3'}
\Vert\ti {\cal H}(\theta) X^0\Vert_{H_1^{-1}}\le C
\Vert X^0\Vert_{H^1_1},\,\,\,\, X^0\in H_1^1.
\ee
where the constant $C$ does not depend on 
$\theta\in K^d$. Hence,
\beqn\la{1212}
\Vert\Om(\theta)X^0\Vert^2_{H_1^{0}}
=(X^0,\ti {\cal H}(\theta) X^0)\le 
\Vert X^0\Vert_{H_1^{1}}
\Vert\ti {\cal H}(\theta) X^0\Vert_{H_1^{-1}}
\le C\Vert X^0\Vert^2_{H^1_1}.
\eeqn
Since, $\Om(\theta)=\Om^*(\theta)$,
the bound (\ref{1212}) implies (\ref{estOm}).

(ii) Condition {\bf R2} implies that
$$
\Vert X^0\Vert_{H^1_1}
\Vert\ti {\cal H}(\theta)X^0\Vert_{H^{-1}_1}\ge
( X^0,\ti{\cal H}(\theta)X^0)
\ge\kappa^2\Vert X^0\Vert^2_{H^1_1}.
$$
Hence,
$\Vert\ti {\cal H}(\theta)X^0\Vert_{H^{-1}_1}
\ge\kappa^2\Vert X^0\Vert_{H^1_1}.$
Therefore,
$\Vert \ti{\cal H}^{-1}(\theta) X^0\Vert_{H_1^{1}}
\le \kappa^{-2}\Vert X^0\Vert_{H_1^{-1}}.$
In particular, (\ref{estH-1}) follows. \bo
\begin{remark}\la{rR'}
Condition {\bf R2'} implies condition {\bf R2}.
\end{remark}
{\bf Proof}.
 Indeed, for $X^0=(\varphi(y),u)\in H^1_1$ and $\theta\in K^d$
we have 
\beqn\la{esopH}
( X^0,\ti{\cal H}(\theta)X^0)
&=& \int\limits_{T_1^d}
\Big[\overline{\varphi}(y)
[(i\nabla_y+\theta)^2+m_0^2]\varphi(y)
+\overline{\ti R_\Pi(\theta,y)}\cdot \overline{ u(\theta)}
\varphi(y)\nonumber\\
&&+\ti R_\Pi(\theta,y)\cdot  u(\theta)
\overline{\varphi}(y)\Big]\,dy
+\om^2_*(\theta)| u(\theta)|^2
\nonumber\\
&=&\int\limits_{T_1^d}
\left[|(i\nabla_y+\theta)\varphi(y)|^2+
\frac{m_0^{2}}{2}
\left|\varphi(y)+\frac{2}{m_0^{2}}\ti R_\Pi(\theta,y) u (\theta)\right|^2\right]\,dy
\nonumber\\
&&+\om^2_*(\theta)| u(\theta)|^2
-\frac{2}{m_0^{2}}\int\limits_{T_1^d}|\ti R_\Pi(\theta,y) u(\theta)|^2\,dy+\frac{m_0^{2}}{2}
\int\limits_{T_1^d}|\varphi(y)|^2\,dy\nonumber\\
&\ge&
\al\int\limits_{T_1^d}|\nabla_y\varphi(y)|^2\,dy
+\left(\frac{m_0^{2}}{2}-\beta d2\pi\right)
\int\limits_{T_1^d}|\varphi(y)|^2\,dy
\nonumber\\
&&+
| u(\theta)|^2\Big(\nu^2_0-\frac{2}{m_0^{2}}\int\limits_{T_1^d}|\ti R_\Pi(\theta,y)|^2\,dy\Big)\nonumber
\eeqn
for some $\beta>0$ and $\al\in(0,\beta/(\beta+1))$.
Take $\beta<m^2_0/(4\pi d)$.
It remains to prove that
$$
\nu^2_0-\frac{2}{m_0^{2}}\ds\int\limits_{T_1^d}|\ti R_\Pi(\theta,y)|^2\,dy>0.
$$
With regard to condition {\bf R2'},
The Parseval equality implies
\beqn\la{6.4}
\int\limits_{T_1^d}|\ti R_\Pi(\theta,y)|^2dy
=\int\limits_{T_1^d}\Big|
\sum\limits_{k\in\Z^d}e^{ik\theta}R(k+y)\Big|^2\,dy
\le \int\limits_{T_1^d}\Big|
\sum\limits_{k\in\Z^d}|R(k+y)|\Big|^2\,dy
<\nu_0^2 m_0^2/2.\nonumber\bo
\eeqn

\subsection{Existence of the  Schr\"odinger group}
Recall that $\om_l(\theta)>0$ ($F_l(\theta,\cdot)$), 
$l=1,2,\dots,$ are the eigenvalues
(orthonormal eigenvectors)
 of the operator $\Om(\theta)$ in $H_1^0$.
Let us prove  the existence of solutions to the Cauchy problem 
(\ref{CPF'}).
We represent $X^0(\theta,t)$ in the form
\be\la{9.11'}
X^0(\theta,t)=\sum\limits_{l=1}^\infty
A_l(t)F_l(\theta,r),\,\,\,\,t\in\R,
\ee
where  $A_l(t)\equiv A_l(\theta,t)$ is the unique solution
of the Cauchy problem
$$
\ddot A_l(t)=-\om_l^2(\theta)A_l(t),\,\,\,\,(A_l(t),\dot A_l(t))|_{t=0}=(A^0_{0l},A^1_{0l}),
$$
and $A^i_{0l}\equiv A^i_{0l}(\theta)=(F_l(\theta,\cdot),X^i_{0}(\cdot))$,
$i=0,1$.
Hence,
\be\la{9.11''}
A_l(t)=\cos\om_l(\theta) t A^0_{0l}
+\frac{\sin\om_l(\theta) t}{\om_l(\theta)}A^1_{0l}.
\ee
By the energy conservation, this yields
$$
\frac{|\dot A_l(t)|^2}{2}+\om^2_l(\theta)\frac{ |A_l(t)|^2}{2}=
\frac{|A^1_{0l}|^2}{2}+\om^2_l(\theta)\frac{| A^0_{0l}|^2}{2}.
$$
Summing up, for $t\in\R$, we obtain  
\be\la{9.14'}
\frac{1}{2}\Vert \dot{X^0}(\theta,t)\Vert^2_{H^0_1}
+\frac{1}{2}(X^0(\theta,t),\ti{\cal H}(\theta)X^0(\theta,t))=
\frac{1}{2}\Vert X_0^1\Vert^2_{H^0_1}
+\frac{1}{2}(X_0^0,\ti{\cal H}(\theta)X_0^0)\le
C\Vert X_0\Vert^2_{\HH^1}
\ee
by (\ref{10.3'}). Hence, the solution (\ref{9.11'})
exists and is unique.

Further, relation (\ref{9.11''}) implies (\ref{rep}).
Finally, the solution to  problem (\ref{CPF'})
exists; it is unique and can be represented by (\ref{9.12'}).
The bound (\ref{10.11}) follows from (\ref{9.14'})
and (\ref{lowerbound}).
 \bo
\medskip

Now the exponential $\exp\Big(\ti{\cal A}(\theta)t\Big)$
is defined for any chosen value $\theta\in K^d$,
and this exponential is a continuous operator in $\HH^1$.

\subsection{Smoothness of the Schr\"odinger group}
To complete the proof of Proposition \ref{c.a2}, 
 we must prove the smoothness of the exponential
with respect to $\theta$.
This is needed to define the product (\ref{solFtr})
of the exponential and the distribution  $\ti Y_{0\Pi}(\cdot)$.

Consider the operators 
$\exp\Big(\ti{\cal A}'(\theta)t\Big)$,
$t\in\R$, on $\HH^{-1}:=(\HH^1)^*=H^{-1}_1\oplus H^0_1$,
where $\ti{\cal A}'(\theta)$ is the  formal adjoint operator
to $\ti{\cal A}(\theta)$:
$$
(X,\ti{\cal A}'(\theta)Z)_{H_1^0}=
(\ti{\cal A}(\theta)X,Z)_{H_1^0},\,\,\,X,Z\in C_0^\infty(T_1^d)\times \C^n.
$$
Note that
\be\la{tiA'}
\ti{\cal A}'(\theta)=\ti{\cal A}^T(\theta)=\left(\ba{ll}
0&-\ti {\cal H}(\theta)\\
1&0\ea\right).
\ee
\begin{lemma}\la{l9.7}
For  any $\al\ge0$, the following bound holds:
\be\la{10.17}
\sup_{|t|\le T}\sup_{\theta\in K^d}
\sum\limits_{|\gamma|\le\al}
\Vert{\cal D}_{\theta}^\gamma
e^{\ti{\cal A}'(\theta) t}X_0\Vert_{\HH^{-1}}\le
C(T)\Vert X_0\Vert_{\HH^{-1}}.
\ee 
\end{lemma}
{\bf Proof.}
For $\al=0$, the bound 
\be\la{9.21'}
\Vert e^{\ti{\cal A}'(\theta) t}X_0\Vert_{\HH^{-1}}\le
C\Vert X_0\Vert_{\HH^{-1}},
\ee
follows from the bound (\ref{10.11}) by duality arguments.
Consider the case $\al=1$. Introduce the  function
$X_\gamma(t):={\cal D}_{\theta}^\gamma
X(\theta,t)$, where
$X(\theta,t)=e^{\ti{\cal A}'(\theta) t}X_0$.
Then
\beqn
\dot X_\gamma(t)=
\ti{\cal A}'(\theta) X_\gamma(t)+
[ {\cal D}_{\theta}^\gamma \ti{\cal A}'(\theta)]
X(\theta,t),\,\,\,\,  X_\gamma(0)=0.\nonumber
\eeqn
Hence,
\beqn
X_\gamma(t)=\int\limits_0^t 
e^{\ti{\cal A}'(\theta)(t-s)}
[ {\cal D}_{\theta}^\gamma \ti{\cal A}'(\theta)]
X(\theta,s)\,ds.
\nonumber
\eeqn
Therefore, by the bound (\ref{9.21'}),
\beqn\la{10.14}
\Vert X_\gamma(t)\Vert_{\HH^{-1}}&\le&
\int\limits_0^t
\Vert e^{\ti{\cal A}'(\theta)(t-s)}
[ {\cal D}_{\theta}^\gamma \ti{\cal A}'(\theta)]
X(\theta,s)\Vert_{\HH^{-1}}\,ds
\nonumber\\
&\le&C\int\limits_0^t
\Vert
[ {\cal D}_{\theta}^\gamma \ti{\cal A}'(\theta)]
X(\theta,s)\Vert_{\HH^{-1}}\,ds.
\eeqn
It follows from (\ref{tiA'}) that
\beqn\nonumber
[{\cal D}_{\theta}^\gamma
\ti{\cal A}'(\theta)]=\left(\ba{ll}
0&-[{\cal D}_{\theta}^\gamma\ti {\cal H}(\theta)]\\
0&0\ea\right),\,\,\,\,
[{\cal D}_{\theta}^\gamma\ti {\cal H}(\theta)]
:=\left(
\ba{cc}
2({\cal D}_{\theta}^\gamma \theta) (i\nabla_y+\theta)
 & [{\cal D}_{\theta}^\gamma\ti S(\theta)]\\
 \,[{\cal D}_{\theta}^\gamma\ti S^*(\theta)]\, 
& 2\omega_*(\theta){\cal D}_{\theta}^\gamma\omega_*(\theta)
\ea\right).
\eeqn
Here $[{\cal D}_{\theta}^\gamma\ti S(\theta)]\,
u:=[{\cal D}_{\theta}^\gamma\ti R_\Pi(\theta,\cdot)]
u$, $u\in\C^n$, and
$[{\cal D}_{\theta}^\gamma\ti S^*(\theta)]\,
\varphi(y):=
\ds\int[{\cal D}_{\theta}^\gamma\ti R_\Pi(-\theta,y)]
\varphi(y)\,dy$.
Hence, if $\theta\in K^d$, then
\beqn\la{10.15}
\Vert 
[ {\cal D}_{\theta}^\gamma \ti{\cal A}'(\theta)]
X(\theta,s)\Vert_{\HH^{-1}}&=&
\Vert [ {\cal D}_{\theta}^\gamma \ti{\cal H}(\theta)]
X^1(\theta,s)\Vert_{H_1^{-1}}\le
C\Vert X^1(\theta,s)\Vert_{H_1^{0}}
\nonumber\\
&\le&
C\Vert e^{\ti{\cal A}'(\theta)s}
X_0\Vert_{\HH^{-1}}
\le C\Vert X_0\Vert_{\HH^{-1}},
\eeqn
by the bound (\ref{9.21'}).
Inequalities  (\ref{10.14}) and (\ref{10.15})
imply the bound (\ref{10.17}) with $\al=1$.
 For $\al>1$, the estimate follows by induction.
\hfill$\bo$
\subsection{Dual group}
Here we complete the proof of  the bound (\ref{boundsol})
by  duality arguments.
Introduce the dual space ${\cal E}':=H^{-1,-\al}(\R^d)\oplus
L^{-\al}\oplus H^{0,-\al}(\R^d)\oplus L^{-\al}$
of functions $Z$ with  finite norm
\beqn\nonumber
(\bre Z\bre'_{0,-\al})^2:=
\Vert \psi\Vert_{-1,-\al}^2+
\Vert\pi\Vert_{0,-\al}^2+\Vert u\Vert_{-\al}^2+
\Vert v\Vert_{-\al}^2.
\eeqn
For  $Z\in {\cal E}'$,
we have $\ti Z_\Pi(\theta,\cdot)\in H^\al(K^d;\HH^{-1})$.
\begin{lemma}\la{c9.9}
Let $\al$ be  even and let $\al\le -2$.
Then
\be\la{9.25'}
\sup_{|t|\le T }\bre W'(t)Z\bre'_{0,-\al}
\le C(T)\bre Z\bre'_{0,-\al}.
\ee
\end{lemma}
{\bf Proof.}
Note first that  
\be\la{2.15}
(\bre Z\bre'_{0,-\al})^2\sim
\sum_{|\gamma|\le-\al}
\int\limits_{K^d}
\Vert {\cal D}_\theta^\gamma
\ti Z_\Pi(\theta,\cdot)
\Vert^2_{\HH^{-1}}\,d\theta
\ee
for $Z\in {\cal E}'$. Indeed, 
\beqn
\Vert u\Vert_{-\al}^2\!\!\!&=&\!\!\sum\limits_{k\in\Z^d}
\langle k\rangle^{-2\al}|u(k)|^2=
C\int\limits_{T^d}|(1-\De_\theta)^{-\al/2}\ti u(\theta)|^2\,d\theta,\nonumber\\
\Vert \psi\Vert_{-1,-\al}^2\!\!\!&=&\!\!
\Vert \langle x\rangle^{-\al}\Lambda^{-1}\psi(x)\Vert^2_{L^2}\sim \Vert 
 \Lambda^{-1}\langle x\rangle^{-\al}\psi(x)\Vert^2_{L^2}=C\Vert 
 (1+|\xi|^2)^{-1/2}
(1\!-\!\De_\xi)^{-\al/2}\hat\psi(\xi)\Vert^2_{L^2}
\nonumber\\
&\sim&
\sum\limits_{m\in\Z^d}\int\limits_{K^d}
(1+|2\pi m+\theta|^2)^{-1/2}|(1-\De_\theta)^{-\al/2}
\hat \psi(2\pi m+\theta)|^2\,d\theta
\nonumber\\
&\sim&
\sum\limits_{m\in\Z^d}\int\limits_{K^d}
(1+|m|^2)^{-1/2}|(1-\De_\theta)^{-\al/2}
\hat \psi(2\pi m+\theta)|^2\,d\theta
\nonumber\\
&\sim&
\int\limits_{K^d}
\Vert(1-\De_\theta)^{-\al/2}
\ti \psi_\Pi(\theta,\cdot)\Vert^2_{H^{-1}(T_1^d)}\,d\theta.
\eeqn
Hence, by Lemma \ref{l9.7} and by (\ref{2.15}),
\beqn\nonumber
(\bre W'(t)Z\bre'_{0,-\al})^2&\sim&
\sum_{|\gamma|\le-\al}
\int\limits_{K^d}
\Vert {\cal D}_\theta^\gamma
\Big(e^{\ti{\cal A}'(\theta)t}\ti Z_\Pi(\theta,\cdot)\Big)
\Vert^2_{\HH^{-1}}\,d\theta\nonumber\\
&\le C(t)&
\sum_{|\gamma|\le-\al}\int\limits_{K^d}
\Vert {\cal D}_\theta^\gamma
\ti Z_\Pi(\theta,\cdot)
\Vert^2_{\HH^{-1}}\,d\theta
\sim C(t)(\bre Z\bre'_{0,-\al})^2.\,\,\,
\nonumber \bo
\eeqn
\begin{cor}\la{adjproblem}
The bound (\ref{boundsol}) follows from 
(\ref{9.25'}) by the duality considerations.
\end{cor}

\setcounter{equation}{0}
 \section{ Appendix B: Crossing points}
\subsection{Proof of  Lemmas \re{lc*} and \re{lc}}
Let us prove  Lemma \re{lc*}.
For any chosen $\theta\in \R^d$, the Schr\"odinger operator
$\ti{\cal H}(\theta)$
admits the  spectral resolution
$$
\ti{\cal H}(\theta)=
\sum\limits_{l=1}^{\infty}\lambda_l(\theta)P_l(\theta),
$$
where
$0<\lambda_1(\theta)\le\lambda_2(\theta)\le\dots$,
and $P_l(\theta)$
are one-dimensional orthogonal projectors in $H_1^0$.
Further, let us take an arbitrary point $\Theta\in \R^d$
and a number 
$\Lambda\in (\lambda_M(\Theta),\lambda_{M+1}(\Theta))$, where $M\ge N$ and $\lambda_M(\Theta)<\lambda_{M+1}(\Theta)$. Then
$\Lambda\not=\lambda_l(\theta)$ for $\theta\in {\cal O}(\Theta)$ 
if ${\cal O}(\Theta)$  is a sufficiently small neighborhood of $\Theta$.
Write
\beqn\nonumber
\left. \ba{ccl}
\ti{\cal H}^\Lambda(\theta)&=&
\sum\limits_{\lambda_l(\theta)<\Lambda}\lambda_l(\theta)P_l(\theta)\\
P^\Lambda(\theta) &=&
\sum\limits_{\lambda_l(\theta)<\Lambda}P_l(\theta)
\ea\right|\,\,\,\,\theta\in {\cal O}(\Theta).
\eeqn
Further, let us choose a contour $\Gamma_\Lambda$
(in the complex plane $\C$)
 surrounding the interval $(0,\Lambda)$
such that $\Lambda\in \Gamma_\Lambda$.
In this case, by the Cauchy theorem,
\beqn\nonumber
\left.
\ba{ccl}
\ti{\cal H}^\Lambda(\theta)&=&
\ds\int\limits_{\Gamma_\Lambda}\frac{\lambda d\lambda}{\ti {\cal H}(\theta)-\lambda}\\
P^\Lambda(\theta) &=&
\ds\int\limits_{\Gamma_\Lambda}
\frac{ d\lambda}{\ti {\cal H}(\theta)-\lambda}
\ea\right|\,\,\,\,\theta\in {\cal O}(\Theta).
\eeqn
Finally, by (\ref{H(theta)})-(\ref{2.6'}) and 
by condition {\bf R1},
the mapping $\ti {\cal H}(\theta)$ is an analytic operator-valued
function in $\theta\in {\cal O}_c(\Theta)$,
where ${\cal O}_c(\Theta)$ is a  complex neighborhood of ${\cal O}(\Theta)$.
Therefore, the same integrals converge for
$\theta\in {\cal O}_c(\Theta)$, and the functions
$P^\Lambda(\theta)$ and $\ti{\cal H}^\Lambda(\theta)$
are analytic in a smaller neighborhood ${\cal O}'_c(\Theta)$.
Reducing ${\cal O}'_c(\Theta)$ again, we can choose a basis
$e_1(\theta),\dots,e_M(\theta)$ in the space
$R^\Lambda(\theta):=P^\Lambda(\theta)H_1^0$, where 
the functions $e_l(\theta)$
depend analytically on $\theta\in {\cal O}_c(\Theta)$.
For example, it suffices to choose an arbitrary basis
$e_1(\Theta),\dots,e_M(\Theta)$ and set
$e_l(\theta)=P^\Lambda(\theta)e_l(\Theta)$.
The operator $\ti{\cal H}(\theta)$ on the invariant space
$R^\Lambda(\theta)$ can be identified with the corresponding matrix
\be\la{10.1}
\ti{\cal H}^\Lambda(\theta)=
\Big(\ti{\cal H}_{kl}(\theta)\Big)_{k,l=0,\dots,M},
\ee
which depends analytically on $\theta\in{\cal O}_c(\Theta)$.
Therefore, the eigenvalues
$\lambda_1(\theta),\dots,\lambda_M(\theta)$
and eigenvectors $F_1(\theta),\dots,F_M(\theta)$
of this matrix
can be chosen as real-analytic functions of
$\theta\in{\cal O}_r(\Theta)\setminus C_*^\Lambda$,
where ${\cal O}_r(\Theta):={\cal O}_c(\Theta)\cap \R^d$
and $C_*^\Lambda$ is a subset of $\R^d$ of Lebesgue measure zero.
This can be proved by using the methods of \ci[Appendix]{DKS1}.
It remains to pass to the limit as $\Lambda\to \infty$ and define
$C_*:=\cup_1^\infty C_*^\Lambda$.
Finally, $\om_l(\theta):=\sqrt{ \lambda_l(\theta)}$.
After this, relations (\ref{enum}) and (\ref{enum'}) follow as in
\ci[Appendix]{DKS1}.
\hfill$\bo$

 Lemma \re{lc} can be proved in a similar way.

\subsection{Proof of Lemma 
\re{l45}}
First let us show that  conditions {\bf E1}, {\bf E2}
hold for $R_C(x)\equiv 0$ corresponding to  $C_1=\dots=C_N=0$.
Indeed, in this case, relation (\ref{H(theta)}) becomes
$$
\ti {\cal H}(\theta):=\left(
\ba{cc}
(i\nabla_y+\theta)^2+m_0^2 & 0\\
0 & \omega_*^2(\theta)
\ea\right).
$$
Therefore, $\om_l(\theta)$ are equal to either $\om_*(\theta)$
or $\sqrt{(2\pi k+\theta)^2+m_0^2}$, $k\in \Z^d$.
Namely, $\om_*(\theta)$ corresponds to the eigenvectors $(0,u)$
with an arbitrary $u\in\R^n$. The square root corresponds to the 
eigenvectors
$F_k(\theta,y)=(e^{-2\pi i k\cdot y},0)$ with $k\in \Z^d$.
It can be readily be seen that  conditions {\bf E1} and {\bf E2}
hold in this case.

Further, choose  an arbitrary $l=1,2,\dots$,
a point $\Theta\in\R^d\setminus {\cal C}_*$
and a bound
$\Lambda \in(\lambda_M(\theta),\lambda_{M+1}(\theta))$
as above (with $M\ge l$). The function $R_C(x)$ and the
corresponding operator  $\ti {\cal H}^\Lambda_C(\theta)$ 
depend analytically on $(\theta,C)\in\C^d\times \C^N$.
Moreover, $R_C(x)$ satisfies conditions {\bf R1} and {\bf R2'} for $C\in B_\ve$
with a sufficiently small $\ve>0$.
 Therefore, as in the proof of Lemma \ref{lc*},
 the corresponding eigenvalues $\om_l(\theta,C)$, $l=1,\dots, M$,
are also  analytic functions of $(\theta,C)$
in the  domain ${\cal M}_l(\Theta)=
{\cal O}_c\setminus {\cal C}$,
where ${\cal O}_c$ is a complex neighborhood of 
${\cal O}_r(\Theta)\times B_\ve$,
and ${\cal C}$ is a proper analytic subset
of ${\cal O}_c$. Hence, the corresponding
determinant $D_l(\theta, C)$
is an analytic function of ${\cal M}_l(\Theta)$.
Further,  ${\cal M}_l(\Theta)$ is an open connected set since
${\cal C}$ is a proper analytic subset. Therefore,
$D_l(\theta,C)\not\equiv 0$ on ${\cal M}_l(\Theta)$ since
$D_l(\theta,0)\not\equiv 0$,
$\theta\in \R^d$.
Further, introduce  the set
$$
M_{1l}=\{C\in B_\ve: D_l(\theta,C)\not\equiv0 \}.
$$
The set $B_\ve\setminus M_{1l}$
cannot contain
any open ball, since otherwise $D_l(\theta,C)\equiv0$.
Hence, $M_{1l}$ is an open dense set in $B_\ve$.
It remains to note that $M_1=\cap_l M_{1l}$ is thus
a dense subset of $B_\ve$.
For $M_2$, the proof is similar.
\hfill$\bo$

\setcounter{equation}{0}
 \section{ Appendix C: Covariance in the spectral representation}

Introduce the matrix-valued operator
\be\la{G(t,theta)}
G(\theta,t):=
e^{\ti{\cal A}(\theta)t}=
\left(\ba{cc}
\cos\Om(\theta)t&\sin\Om(\theta)t\,\Om^{-1}(\theta)\\
-\Om(\theta)\sin\Om(\theta)t&\cos\Om (\theta)t
\ea\right).
\ee
Note that  we can represent the matrix $G(\theta,t)$
in the form
\be\la{repGt}
G(\theta,t)=\cos\Om(\theta)t\, I+\sin\Om(\theta)t\, C(\theta),
\ee
where $I$ stands for  the  unit matrix, and
\beqn\nonumber
C(\theta):=\left(\ba{cc}
0&\Om^{-1}(\theta)\\
-\Om(\theta)&0
\ea\right).
\eeqn
In this case, the solution of (\ref{CPF}) has the  form
$
{\ti Y}_\Pi(\theta,r,t)=G(\theta,t){\ti Y}_{0\Pi}(\theta,r)$,
$r\in {\cal R}$.
Using (\ref{repGt})  
and (\ref{qdpFt}), we obtain
\beqn\la{ap1}
\ti Q_t(\theta,r,\theta',r')
&=&
E[{\ti Y}_\Pi(\theta,r,t)\otimes
\overline{{\ti Y}_\Pi(\theta',r',t)}]
\nonumber\\
&=&\cos\Om(\theta) t~\ti Q_0(\theta,r,\theta',r')\, \cos\Om(\theta')t\nonumber\\
&&+
\sin\Om(\theta)t~ C(\theta)\ti Q_0(\theta,r,\theta',r')
 C^T(\theta')\, \sin\Om(\theta') t
\nonumber\\
&&+\cos\Om(\theta) t~\ti Q_0(\theta,r,\theta',r')
 C^T(\theta')\, \sin\Om(\theta') t
\nonumber\\
&&+
\sin\Om(\theta) t~ C(\theta)
\ti Q_0(\theta,r,\theta',r')\,\cos\Om(\theta')t.
\eeqn
By (\ref{ap22}), 
we see that 
\beqn\la{ap1'}
\ti q_t(\theta)
&=&G(\theta,t)\ti q_0(\theta)G^*(\theta,t)
=\cos\Om(\theta) t~\ti q_0(\theta)\, \cos\Om(\theta)t
\nonumber\\
&&+\cos\Om(\theta) t~\ti q_0(\theta)
 C^T(\theta)\, \sin\Om(\theta) t+
\sin\Om(\theta) t~ C(\theta)
\ti q_0(\theta)\,\cos\Om(\theta)t
\nonumber\\
&&+
\sin\Om(\theta)t~ C(\theta)\ti q_0(\theta)
 C^T(\theta)\, \sin\Om(\theta) t,
\eeqn
where $\ti q_t(\theta)$
is the integral operator with the kernel
 $\ti q_t(\theta,r,r')$ defined by (\ref{qt}).

For the simplicity of our manipulations, we
 assume now that
the set of  ``crossing'' points $\theta_*$
is empty, i.e., $\om_l(\theta)\ne\om_{l'}(\theta)$
for any $l,l'\in\N$, and the functions
$\om_l(\theta)$ and $F_l(\theta,r)$ are real-analytic.
(Otherwise we need a partition of unity
(\ref{part})).
Consider the first term in the RHS of (\ref{ap1'})
and represent it in the form
\beqn
&&\cos\Om(\theta) t~\ti q_0(\theta)\, \cos\Om(\theta)t
=\Big(\sum_{l,l'}F_l(\theta,r)\Big(\cos\om_l(\theta)t~
p_{ll'}(\theta)\cos\om_{l'}(\theta)t
\Big)\overline{F_{l'}}(\theta,r')
\nonumber\\
&&=
\sum_{l,l'}F_l(\theta,r) \frac{1}{2}
\Big[\cos(\om_l(\theta)\!-\!\om_{l'}(\theta))t +
\cos(\om_l(\theta)\!+\!\om_{l'}(\theta))t\Big] p_{ll'}(\theta)\overline{F_{l'}}(\theta,r'),\,
\eeqn
where $p_{ll'}(\theta)=
\Big(p^{ij}_{ll'}(\theta)\Big)_{i,j=0}^1
=\Big(F_l(\theta,\cdot),(\ti q_0^{ij}(\theta)F_{l'})(\theta,\cdot)
\Big)_{i,j=0}^1$
($p^{ij}_{ll'}(\theta)$ are introduced
in (\ref{pij})).
Similarly, we can rewrite
the remaining three terms in the RHS of (\ref{ap1'}).
Finally,
\beqn\la{6.10}
\ti q^{ij}_t(\theta,r,r'):=\sum_{l,l'=1}^{\infty}
F_l(\theta,r)r^{ij}_{ll'}(t,\theta) \otimes\overline{F_{l'}}(\theta,r'),
\eeqn
where  $r_{ll'}(t,\theta)=(r^{ij}_{ll'}(t,\theta))_{ij=0}^1$ 
are defined in (\ref{ap.2}). 


\end{document}